\documentclass{jfm}
\usepackage{graphicx}

\usepackage{natbib}
\newcommand{\bm}[1]{\mbox{\boldmath $#1$}}

\newcommand\Rey{\mbox{\it{Re}}}  
\newcommand\De{\mbox{\it{De}}}
\newcommand\ttil{\tilde{t}}
\newcommand\xtil{\tilde{x}}
\newcommand\detil{\tilde{\partial}}
\newcommand{\eqref}[1]{(\ref{#1})}

\newcommand\etal{\mbox{\textit{et al.}}}

\newcommand\eg{e.g.}
\newcommand\ie{i.e.}

\title[Nonlinear dynamics of the viscoelastic Kolmogorov flow]{Nonlinear dynamics of the viscoelastic Kolmogorov flow}

\author[A.~Bistagnino, G.~Boffetta, A.~Celani, A.~Mazzino, A.~Puliafito, M. ~Vergassola]%
{\mbox{A. \ns B \ls I \ls S \ls T \ls A \ls G \ls N \ls I \ls N \ls O$^1$},
\ns \mbox{G. \ns B \ls O \ls F \ls F \ls E \ls T \ls T \ls A$^1$},
\ns \mbox{A. \ns C\ls E\ls L\ls A\ls N\ls I$^2$},
\ns \mbox{A. \ns M\ls A\ls Z\ls Z\ls I\ls N\ls O$^{3,5}$},
\ns  \mbox{A. \ns P\ls U\ls L\ls I\ls A\ls F\ls I\ls T\ls O$^{2,3}$}
\ns \and \mbox{M. V\ls E\ls R\ls G\ls A\ls S\ls S\ls O\ls L\ls A$^4$}}

\affiliation{
$^1$ Dipartimento di Fisica Generale,
Universit\`a di Torino and
Istituto Nazionale di Fisica Nucleare, Sez. di Torino, V. Giuria 1,
10125 Torino, Italy 
\\[\affilskip]
$^2$ CNRS, INLN, 1361 Route des Lucioles, 06560 Valbonne,
France
\\[\affilskip]
$^3$ Dipartimento di Fisica, 
Universit\`a di Genova, Via Dodecaneso
33, 16146 Genova,  Italy
\\[\affilskip]
$^4$ CNRS URA 2171, Inst. Pasteur, 25 rue du Dr. Roux, 75724 Paris 
Cedex 15, France.
\\[\affilskip]
$^5$ Istituto Nazionale di Fisica Nucleare, Sez. di Genova, Via Dodecaneso
33, I--16146 Genova, Italy
}

\begin{document}

\maketitle

\begin{abstract}
The weakly nonlinear regime of a viscoelastic Navier--Stokes fluid is
investigated. For the purely hydrodynamic case, it is known that
large-scale perturbations tend to the minima of a Ginzburg-Landau
free-energy functional with a double-well (fourth-order) potential.
The dynamics of the relaxation process is ruled by a one-dimensional
Cahn--Hilliard equation that dictates the hyperbolic tangent profiles
of kink-antikink structures and their mutual interactions.  For the
viscoelastic case, we found that the dynamics still admits a
formulation in terms of a Ginzburg--Landau free-energy functional. For
sufficiently small elasticities, the phenomenology is very similar to
the purely hydrodynamic case: the free-energy functional is still a
fourth-order potential and slightly perturbed kink-antikink structures
hold. For sufficiently large elasticities, a critical point sets in:
the fourth-order term changes sign and the next-order nonlinearity
must be taken into account. Despite the double-well structure of the
potential, the one-dimensional nature of the problem makes the
dynamics sensitive to the details of the potential. We analysed the
interactions among these generalized kink-antikink structures,
demonstrating their role in a new, elastic instability. Finally,
consequences for the problem of polymer drag reduction are presented.

\end{abstract}

\section{Introduction}

The derivation of coarse-grained equations of motion, averaging out
microscopic degrees of freedom and retaining only those features
relevant to the process of interest, is a major goal in many different
scientific domains.  A first classical example is the dynamics of
celestial bodies, the physical problem which motivated the
introduction of asymptotic techniques to systematically average over
rapidly rotating, angular degrees of freedom. More recently, many
interesting phenomena in biological contexts (\eg~ related to domain
formation in lipid membrane, bilayer fusion, and cooperative motions
associated with phase changes) have been found to occur on times and
length scales much larger than the typical times and scales where the
classical molecular-dynamics methods are applicable \cite[][]{VK05}.
To reach those larger length-scales, one resorts to coarse-grained
models that employ effective interaction potentials \cite[][]{KVLN04}.

Another relevant example of coarse-grained model comes from
climatology. The current numerical models for prediction of
weather and climate involve general circulation models. They consist
of coupled, nonlinear partial differential equations, discretized in
space and time for the purpose of numerical simulations.  The current
generation of supercomputers supports mesh spacing of the order of 200
km for short-term climate simulations. However, many important
physical processes occur on smaller scales (\eg~the cloud cover in the
boundary layer) and they significantly affect the large-scale dynamics
of resolved fields.  A powerful way to incorporate the unresolved
dynamics is provided by suitable coarse-grained stochastic models
\cite[][]{KMK03}.

Finally, in the framework of phase-ordering kinetics, the concept of
coarse-grained description plays a crucial role for the
order-parameter dynamics. Coarsening is intimately related to the fact
that domain growth is a scaling phenomenon: domain patterns at
different times differ solely by a global scale factor \cite[see the review by][]{B02}.  A suitable coarse-grained description for systems
where the order parameter is not conserved (\eg~for anti-ferromagnetic
ordering) is provided by the time-dependent Ginzburg--Landau
equation. When the order parameter is conserved, as in phase
separation, the coarse-grained dynamics is ruled by the Cahn--Hilliard
equations \cite[][]{B02}: \begin{equation} \frac{\partial w}{\partial
t} = \partial^2_x \frac{\delta F}{\delta w}\,,
\label{landau} \end{equation} where $w({\bf x},t)$ is a suitable coarse-grained
order-parameter and $F$ a Landau free-energy functional:
\begin{equation} F[w]=\int {\rm d}{\bf x} 
\left[\frac{\lambda}{2}|\nabla w|^2+ I(w)\right] \,.  \label{func}
\end{equation} The potential $I(w)$ typically has a double-well 
structure, whose minima correspond to two equilibrium states. 
$\lambda$ is a positive constant related to the distance between the
equilibrium states and thus the size of the interface between them.

In fluid mechanics, the Cahn--Hilliard equations \eqref{landau} play a
fundamental role in the stability analysis of large-scale
perturbations. In a variety of situations, it turns out that the
evolution of large-scale perturbations is governed by equation \eqref{landau},
with a fourth-order potential $I(w)$
\cite[see][]{N76,S85,P87,MY99}. The structure of the potential
controls the profile and the interactions of the so-called
kink-antikink structures observed in snapshots of the flow
\cite[][]{She87}.

In the present paper, we focus our attention on a simple model of
viscoelastic flows, the so-called viscoelastic Kolmogorov flow. Its
linear stability analysis has been recently investigated by
\cite{B05}, while the turbulent regime and its massive drag
reduction effects have been studied by \cite{BCM05}.  Here, we analyse
the weakly nonlinear dynamics, intermediate between the linear stage of
evolution and the fully turbulent regime.

The starting points of our analysis are three results obtained by
\cite{B05} for the linearized stage: i) The most unstable perturbation
has a long wavelength (large-scale) compared to the period of the
basic Kolmogorov flow; ii) Its linear evolution is captured by
asymptotic multiscale methods, at least up to moderate elasticities of
the flow; iii) The most unstable perturbation is transverse with
respect to the basic flow. 

Multiscale asymptotic methods can be applied, as in the Newtonian
case, to show that the evolution in the presence of polymers obeys a
one-dimensional Cahn--Hilliard equation of the form
\eqref{landau}. The point demonstrated here is that there exists a
critical value of the elasticity, where the potential $I(w)$ passes
from the fourth to the sixth order in the field $w$. This corresponds
to a triple critical point. Due to the one-dimensional character of
the nonlinear dynamics, the transition strongly impinges on the
dynamics of the large-scale perturbation.

Above the critical elasticity, ``hydrodynamic'' kink-antikink
structures are replaced by generalized kinks and anti-kinks and their
annihilation processes are shown to be severely slowed down. Moreover,
below the critical value of the elasticity, the mechanism of
instability is linear and nonlinear terms stabilize the
flow. Conversely, above the critical value, we show that a
sub-critical, nonlinear mechanism of instability takes place, provided
the initial amplitude of the perturbation be sufficiently strong.

The paper is organized as follows. In \S\,\ref{sec:viscoNS} and
\ref{sec:linearanalysis}, we describe the viscoelastic model
considered in the sequel and briefly review the results by \cite{B05}
needed here. In \S\,\ref{sec:standardCH}, we use multiscale methods to
derive the coarse-grained equations for the perturbations. In
\S\,\ref{sec:generalisedCH}, we study the system around the triple
critical point and work out the evolution equations in its
neighborhood. In \S\,\ref{sec:variationalf} and \ref{sec:numres}, we
reformulate the asymptotic behaviour of the coarse-grained equations
in terms of variational analysis and present the numerical results
that corroborate our analytical predictions. Finally, in
\S\,\ref{sec:dragred} we address the problem of drag reduction and show
that, even for the weakly unstable regime considered here, the
injection of polymers induces an enhancement of the mean flow
amplitude.

\section{The viscoelastic Navier--Stokes equations}\label{sec:viscoNS}

Several models have been introduced \cite[see \eg][]{H77} to describe
viscoelastic fluids. A powerful class describes the fluid as
non-Newtonian, accounting for the reaction of the polymers onto the
flow via an extra-term in the stress tensor. A popular and
often employed model within this class is the Oldroyd-B \cite[][]{O50},
which is the one considered in the sequel. We briefly review it here
for the sake of completeness.

In the Oldroyd-B model, it is assumed that viscoelastic flows can be
treated as a dilute suspension of elastic dumbbells, \ie~identical
pairs of microscopic beads connected by Hookean springs. The flow is
considered ``external'' to the molecule, neglecting the effects of the
finite size of the polymers on the flow. Furthermore, the polymer
concentration is supposed to be uniform and low enough to neglect
polymer-polymer interactions.

The reaction of the dumbbells on the fluid is treated at a mean-field
level and the study of the dynamics is limited to scales much larger
than the inter-polymer distance. The polymer solution is regarded as a
continuous medium, whose reaction on the flow is described by an
elastic contribution ${\bm T}$ to the total stress tensor of the
fluid.  Its value per unit density depends on the free energy of the
molecule and the thermal noise as \cite[see \eg][]{B87}:
\begin{equation} {\bm T}=- n_p \langle \bm{R} \bm{F} \rangle - n_p
k_B \Theta \, \bm{{\sf 1}} \,, \end{equation} where $n_p$ is the
polymer density, $k_B \Theta$ is the energy associated with thermal
noise, $F_i$ is the dumbbell relaxation force and $R_i$ its elongation
vector. The average is taken over the statistics of the thermal
noise. Assuming the force between the beads to be Hookean with
dynamical coefficient $K_0$, the average in the elastic stress reduces
to $\langle\bm{R}\bm{F}\rangle = -K_0 \langle\bm{R} \bm{R}
\rangle$. The latter is proportional to the conformation tensor ${\bm
\sigma}\equiv\langle{\bm R \bm R}\rangle/R_0^2$, where $R_0$ denotes
the equilibrium spring length.  The inclusion of the extra elastic
stress term in the Navier--Stokes equations leads to the following equation
for the viscoelastic flow:
\begin{equation}
  \partial_t {\bm v} + ({\bm v}\cdot{\bm \partial}) {\bm v}
=-{\bm{\partial} p} + \nu \beta {\partial^2} {\bm v} +
\frac{\nu(1-\beta)}{\tau} {\bm \partial}\cdot ({\bm \sigma}
-\bm{\sf{1}})+ {\bm f}\,.  \label{eq:Diffuu} \end{equation} Here,
$\nu$ is the total kinematic viscosity of the solution, while
$\nu\beta$ and $\nu (1-\beta)$ are the separate contributions by the
solvent and the polymers, respectively, and we have introduced the
dimensionless parameter $\beta=\eta_s/(n_p k_B\Theta\tau+\eta_s)$,
$\eta_s$ being the dynamic viscosity of the solvent.  
$\tau$ is a parameter depending on $K_0$ and $R_0$, representing the typical 
relaxation time of the polymers. A more precise definition of $\tau$ and ${\bm
 \sigma}$ comes in the following. 
Throughout the paper, it is understood that $({\bm \partial \bm
v})_{\alpha\beta}\equiv \partial_\alpha v_\beta$ and $\mathrm
{tr}\left({\bm \partial}{\bm v}\right)= {\bm \partial}\cdot{\bm v}=0$.

An equation for the dynamics of the polymer conformation tensor ${\bm
\sigma}$ is needed to close the
system of equations.  Simple physical reasoning by \cite{B87} gives
the following stochastic equation for the separation ${\bm R}$ between
two beads:
\begin{equation} \label{eq:evR} \dot{\bm{R}} = (\bm{R\cdot\partial})
\bm{v} -\frac{1}{2 \tau} \bm{R}+\sqrt{\frac{R_0^2}{\tau}}\bm{\xi}\; .
\end{equation} On the right-hand side, the first term is the
stretching/compression term, originating from the spatial variation of
the flow experienced by $\bm{R}$, and the last one, $\bm\xi$, is a
white-in-time random process mimicking the effect of thermal noise on
the polymers. The second is the relaxation due to the force between
the beads, proportional to the elongation derivative of the dumbbell
free energy $-\partial E / \partial R_i = -\partial (1/2 K_0 R^2) /
\partial R_i$. A quadratic form of the potential, and thus a linear
Hookean force, is an approximation valid for moderate polymer
elongations. The dynamical coefficient $\tau$ is the same as
the one appearing in \eqref{eq:Diffuu}. Considering it constant
amounts to assume that the polymers have only one relaxation
time. Numerical and theoretical studies point out that this hypothesis is
reasonable \cite[][]{GCS05}. Experiments \cite[see][]{L69,V75,NH95} show that
polymers have a spectrum of typical relaxation times, but they also show that interactions with the fluid
mostly depend on the largest one, that is the one we are
retaining.

Multiplying \eqref{eq:evR} by $\bm{R}$ and averaging over the statistics of the thermal noise
${\bm \xi}$, the following evolution equation for the conformation
tensor ${\bm \sigma}=\langle{\bm R \bm R}\rangle/R_0^2$ is
obtained \cite[see][]{B87}:
\begin{equation}
\partial_t {\bm \sigma} + ({\bm v}\cdot{\bm \partial}) {\bm \sigma}
= ({\bm \partial \bm v})^T \cdot {\bm \sigma} + {\bm \sigma} \cdot
({\bm \partial \bm v})
-\frac{1}{\tau} ({\bm \sigma} -
\bm{\sf{1}}) 
\label{eq:Diffusigma}\; .
\end{equation}

Summarizing, the set of equations \eqref{eq:Diffuu} and \eqref{eq:Diffusigma} 
constitutes the Oldroyd-B model that we shall be considering 
in the sequel. 

Our first step in the investigation of the effect of polymers onto the
stability of the flow will be to find out the basic equilibrium state.
The state will then be perturbed and the resulting equations analysed
using multiscale methods.

\subsection{A basic equilibrium state}

Finding analytically the basic equilibrium state for a generic forcing
${\bm f}$ is a hopeless task already for the Navier--Stokes equations
without polymers. The problem is further complicated here by the additional
term in \eqref{eq:Diffuu} and the coupling with \eqref{eq:Diffusigma}.

The problem simplifies for ${\bm f}\equiv(f(z),0,0)$, inducing a
parallel flow ${\bm U} = (U(z),0,0)$, which trivially annihilates the
advective nonlinear term in \eqref{eq:Diffuu}.  A further substantial
simplification comes from the viscoelastic version of Squire's theorem
(see Appendix \ref{app:squire}), stating that, for parallel flows, the
most unstable perturbations are two-dimensional. We shall therefore
restrict to a two-dimensional flow $(u_x,u_z)$, without any lack of
generality \cite[see also][]{B05}. We further assume
$f(z)=F_0\cos(z/L)$, producing the well-known Kolmogorov flow
\cite[][]{AM60} $U(z)\equiv (V\cos(z/L),0)$, where $V=F_0\,L^2/\nu$.
The corresponding conformation tensor at equilibrium is:
\begin{equation} {\bm \sigma}=\left( \begin{array}{cc} 1+2{\tau}^2\,
(\partial_z U)^2 & {\tau}\, \partial_z U \\
 {\tau}\, \partial_z U  & 1
\end{array}
\right) =
\left(
\begin{array}{cc}
1+2{\tau}^2\, \frac{V^2}{L^2}\sin^2{(\frac{z}{L})} & -{\tau}\frac{V}{L}\,
\sin{{(\frac{z}{L})}} \\ -{\tau}\frac{V}{L}\, \sin{{(\frac{z}{L})}} & 1
\end{array}
\right)
\;.
\label{eq:conftensor}
\end{equation}
This choice also allows to precisely define the Reynolds number of the flow as
$\Rey=V L / \nu$. In this model the elasticity of the polymers is taken into
account by the relaxation time $\tau$ only. We can thus introduce an
adimensional parameter, the Deborah number $\De= \tau V / L$, to characterize
the elastic properties of the flow.

\section{Some previous results on the linear stability
analysis}\label{sec:linearanalysis}

It has long been known that the Newtonian Kolmogorov flow becomes
unstable for Reynolds numbers $\Rey > \sqrt{2}$ \cite[][]{MS61}: the
evolution of large-scale perturbations is formally described by an
effective diffusive dynamics and instabilities are associated to the
loss of positive-definiteness of the eddy-viscosity tensor.

In the presence of polymers, performing a multiscale analysis
\cite[][]{BLP78,BOH88} on the linearized Oldroyd-B model, one obtains
an explicit expression for the eddy-viscosity tensor, valid for
sufficiently low elasticity \cite[][]{B05}.  The resulting stability
curve in terms of the Reynolds and the Deborah number is reported in figure~\ref{fig:stab}.

\subsection{Multiscale analysis}

Substituting $\bm{v}=\bm{u}+\bm{w}$ into 
(\ref{eq:Diffuu},\ref{eq:Diffusigma}), the equations for the perturbation
fields read:
\begin{eqnarray} {\bm \partial}\cdot{\bm w} =
0\,, & \label{eq:asym1} \end{eqnarray} \begin{eqnarray} \partial_t
{\bm w} + {\bm \partial}\cdot ({\bm u}{\bm w} + {\bm w}{\bm u} +
\bm{w}\bm{w}) & =-{\bm \partial} q + \nu \beta {\partial^2} {\bm w} +
\nu \,(1-\beta)\, {\tau^{-1}}\, {\bm \partial}\cdot {\bm \zeta}\,,
\label{eq:asym2} \end{eqnarray} \begin{eqnarray} \partial_t {\bm
\zeta} + {\bm \partial} \cdot({\bm u} {\bm \zeta} + {\bm w} {\bm
\sigma} + {\bm w} {\bm \zeta}) = ({\bm \partial \bm u})^T \cdot {\bm
\zeta} + ({\bm \partial \bm w})^T \cdot {\bm \sigma} + ({\bm \partial
\bm w})^T \cdot {\bm \zeta} + \nonumber \\ + {\bm \zeta} \cdot ({\bm
\partial \bm u}) + {\bm \sigma} \cdot ({\bm \partial \bm w}) + {\bm
\zeta} \cdot ({\bm \partial \bm w}) -{\tau^{-1}} {\bm \zeta} \; ,
\label{eq:asym3} \end{eqnarray} where $q$ and $\bm \zeta$ are
the perturbations associated to the
pressure term $p$ and the basic stress tensor $\bm \sigma$. In the
linear stability analysis, the nonlinear terms containing the product
of two perturbation fields are supposed to be negligible \cite[see][]{B05}.

As in the Newtonian case, it is assumed that the first unstable
perturbations have periodicity much larger than that of the
basic flow. The validity of this assumption has already been
investigated by \cite{B05} and is satisfied in the range of parameters
considered here.  

In addition to the usual ``fast'' space/time variables $\bm{x},t$,
describing the basic flow, multiscale techniques introduce ``slow''
variables $\bm{\tilde{x}}=\epsilon \bm{x}$, ${\tilde{t}}=\epsilon^2
t$, to describe the large-scale flow, and prescribe to treat the two
sets as independent. This leads to the expansion of the differential
operators:
\begin{equation} \partial_i \to \partial_i + \epsilon
\tilde{\partial}_i , \qquad \partial_t \to \partial_t + \epsilon^2
\tilde{\partial}_t \,, \label{eq:chain} \end{equation} and of the
fields:
\begin{equation}
\begin{array}{l} {\bm w}={\bm
w}^{(0)}(z,t,\tilde{x},\tilde{z},\tilde{t})+ \epsilon{\bm
w}^{(1)}(z,t,\tilde{x},\tilde{z},\tilde{t})+ \epsilon^2{\bm
w}^{(2)}(z,t,\tilde{x},\tilde{z},\tilde{t})+\ldots\,, \\
q=q^{(0)}(z,t,\tilde{x},\tilde{z},\tilde{t})+ \epsilon
q^{(1)}(z,t,\tilde{x},\tilde{z},\tilde{t})+ \epsilon^2
q^{(2)}(z,t,\tilde{x},\tilde{z},\tilde{t})+\ldots\,, \\ {\bm
\zeta}={\bm \zeta}^{(0)}(z,t,\tilde{x},\tilde{z},\tilde{t})
+\epsilon{\bm \zeta}^{(1)}(z,t,\tilde{x},\tilde{z},\tilde{t})+
\epsilon^2{\bm \zeta}^{(2)}(z,t,\tilde{x},\tilde{z},\tilde{t})+\ldots
\,. \end{array} \label{eq:asym4} \end{equation} All of the
functions have the periodicity of the basic flow and are
independent of $x$.

Inserting \eqref{eq:asym4} into \eqref{eq:asym1}-\eqref{eq:asym3} and
collecting terms of the same order in $\epsilon$, the coarse-grained
equation for a large-scale perturbation is obtained as the solvability
condition at the order $\epsilon^2$. In terms of the stream function
$\Psi$, the perturbation evolves according to the non-isotropic
diffusion equation:
\begin{equation} \label{eq:eddyvisc} \tilde{\partial_t} \tilde{\Delta}
\Psi=\nu_{\alpha\beta}\,
\tilde{\partial}^2_{\alpha}\tilde{\partial}^2_{\beta}\Psi\,,
\end{equation} where the eddy-viscosity
tensor $\nu_{\alpha \beta}$ is not positive-definite for
$\Rey>\sqrt{2}$ (in the absence of polymers).  In general, there exist
critical values of $\Rey$ and $\De$
where perturbations start growing in time.

\begin{figure} \centering 
\includegraphics[angle=0, width=0.8\textwidth]{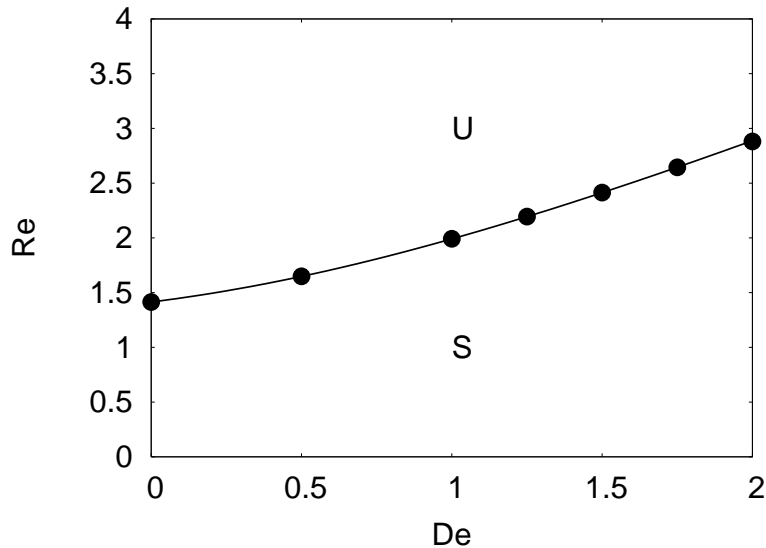}
 \caption{The linear
stability diagram for $\beta=0.769$. Stable and unstable regions are
denoted by S and U, respectively. The bullets represent direct numerical
simulations (DNS) of the complete Oldroyd-B system of equations, confirming
theoretical predictions for this window of parameters.  }
\label{fig:stab} \end{figure}

The phase-space $(\Rey,\De)$ is thus divided in regions where the
eddy-viscosity tensor is positive-definite (the system is linearly stable with
respect to any small perturbation) and where there exists at least one
unstable mode, as shown in figure~\ref{fig:stab} for low Deborah numbers. 
The diagram reveals two kinds of
instabilities. When the Deborah number is sufficiently low, the flow
experiences hydrodynamic-like large-scale transverse instabilities,
captured by multiscale analysis. In this region, the critical Reynolds
number where the flow becomes unstable, grows with $\De$: polymers
stabilize the flow. This has been interpreted by \cite{B05}, and will
be shown in \S\,\ref{sec:dragred}, to be a prelude to the drag reduction effect
observed in the turbulent regime. 

For high values of the Deborah number (not shown in
figure~\ref{fig:stab}) the multiscale analysis predicts the flow to be
unstable, even for very low Reynolds numbers. However, numerical
simulations show that the assumption of scale separation does not hold
and multiscale techniques are not applicable. This region, possibly
characterized by purely elastic instabilities, will not be the concern
of the present investigation which focuses on $0\le\De\le2$. 

If the amplitude of the large-scale perturbation is strong enough
and/or the eddy-viscosity is negative, nonlinear effects are important
and should be taken into account. These two situations correspond to
different scalings of the fields and will be treated separately in the next sections.

\section{Nonlinear dynamics: the standard Cahn--Hilliard
equation}\label{sec:standardCH}

Linear stability analyses, by their very definition, are not able to
capture the full-time dynamics of unstable perturbations: as
perturbations grow in time, their magnitude becomes large and
nonlinearities ought to be taken into account. In the Newtonian case
\cite[][]{S85,She87}, the lowest-order nonlinearity (third-order) is
sufficient to stabilize the linear exponential growth and lead to a
steady state. Here, we show how this occurs and generalize it first to
the viscoelastic case. The next section will then be devoted to the
case where the lowest-order nonlinearity does not stabilize the flow
and higher-order nonlinearities become relevant.

A careful analysis of the linear eddy-viscosity tensor $\nu_{\alpha \beta}$
derived in \eqref{eq:eddyvisc} ensures that for low enough Deborah numbers the
first modes to become linearly unstable are the large-scale transverse
ones. For barely unstable flows we may expect that the perturbations involved
in the nonlinear dynamics will be confined to these modes. This suggests that
the result will be a one-dimensional diffusion equation for the averaged transverse
modes, linearly stable for small-scale modes, involving at least one nonlinear
term.
 
Assume now that the initial amplitude of the large-scale perturbation is
sufficiently small and the system we consider is in the surroundings
of a point of the critical curve. According to
\eqref{eq:eddyvisc}, the average transverse velocity perturbation $ \langle w_z \rangle$ linearly
evolves according to a diffusion equation:
\begin{equation}\label{eq:lapl} \detil_t \langle w_z \rangle \sim -A
\detil^2 \langle w_z \rangle \, ,
\end{equation}
where $A$ is a positive coefficient representing the linear eddy-viscosity
tensor of \eqref{eq:eddyvisc}, restricted to low Deborah numbers (as explained
in the end of \S\,\ref{sec:linearanalysis}). It vanishes on the stability curve, being positive above it and negative below it.

In this equation, all modes are linearly unstable. It needs to be modified to
keep track of the multiscale hypothesis, which requires small-scale modes to
be stable. This is done introducing a bi-Laplacian term into
\eqref{eq:bilapl} to stabilize the small scales (the fourth-order derivative
ensures that this term will be dominant on the small-scale perturbations only):
\begin{equation}\label{eq:bilapl} \detil_t \langle w_z \rangle \sim -A
\detil^2 \langle w_z \rangle -C \detil^4 \langle w_z \rangle \, .
\end{equation}
In general, close to the linear instability threshold, where the
coefficient $A$ changes sign, we do not expect $C$ to vanish. To comply with the stability
requirements, we will request it to have a finite, positive value in the
region of interest.

We expect to find the
presence of a nonlinear term, eventually stabilizing this growth.
This part cannot be played by the advective nonlinearity because of the
one-dimensional character of the equation.
 The next-order nonlinearity is cubic and must contain at least two space
derivatives: one before the whole term, to ensure momentum
conservation, and an additional one to respect space-inversion
symmetries. This yields a nonlinear term: $B\tilde{\partial}_x
\left(\langle w_z^{(1)}\rangle^2\tilde{\partial}_x \langle w_z^{(1)}
\rangle\right)$, where $B$ is some constant related to the (nonlinear) eddy-viscosity.

We can now introduce, as in \S\,\ref{sec:linearanalysis}, the ``slow''
variables $\bm{\tilde{x}}=\epsilon \bm{x}$ and $\tilde{t}$ (notice that we
still do not know the scaling between $t$ and $\tilde{t}$). The space
derivatives must again be expanded as $\partial_i \to \partial_i + \epsilon
\tilde{\partial}_i$. We now have to look for a prescription on how to expand
the different fields in terms of $\epsilon$.

In the vicinity of the marginal eddy-viscosity curve $A \approx 0$, and a Taylor
expansion gives $A \sim \left. \frac{\partial A}{\partial \nu}
\right\arrowvert_{\nu_c} (\nu -\nu_c )$, where $\nu_c$ indicates the
critical viscosity. Balances between the term $A \detil^2 \langle w_z
\rangle$ and both the cubic nonlinearity and $C \detil^4 \langle w_z
\rangle$ yield:
\begin{equation}
\label{AB} B\epsilon^2 \epsilon_w^3 \sim \left.\frac{\partial
A}{\partial \nu}\right\arrowvert_{\nu_c} (\nu -\nu_c ) \epsilon^2
\epsilon_w\,,\qquad \left.\frac{\partial A}{\partial
\nu}\right\arrowvert_{\nu_c} (\nu -\nu_c ) \epsilon^2 \epsilon_w \sim
C \epsilon^4 \epsilon_w \,.\end{equation} Here, $\epsilon_w$ is the
scaling of the amplitude of the velocity perturbation $\bm{w}$.

Equations \eqref{AB} completely define the scaling for the velocity
perturbation and the distance from the critical viscosity:
\begin{equation}\label{XXX} \epsilon=\epsilon_w\,, 
\qquad \frac{\nu - \nu_c}{\nu}
\sim \epsilon^2 \Rightarrow \nu=\nu_c(1-\epsilon^2)\,.
\end{equation}
It follows from (\ref{XXX}) that the Reynolds number
$\Rey=\Rey_c\left(1+\epsilon^2\right)$. The comparison of any of the
previous terms with the (slow) time derivative of $\bm{w}$ gives the
scaling $\tilde{t}=\epsilon^4 t$. As for the scaling of the
polymer conformation tensor, balancing $\zeta / \tau$ and $(\partial
w_z) \cdot \sigma$, we obtain that the scaling of $\bm{\zeta}$
coincides with $\epsilon_w$. The same equality holds for the pressure
field.

Summarizing, the fields are expanded as:
\begin{equation}
\begin{array}{l}
{\bm w}=\epsilon{\bm w}^{(1)}(z,\tilde{x},\tilde{t})+
 \epsilon^2{\bm w}^{(2)}(z,\tilde{x},\tilde{t})+\ldots\,, \\
q=\epsilon q^{(1)}(z,\tilde{x},\tilde{t})+
 \epsilon^2 q^{(2)}(z,\tilde{x},\tilde{t})+\ldots\,, \\
{\bm \zeta}=\epsilon{\bm \zeta}^{(1)}(z,\tilde{x},\tilde{t})+
\epsilon^2{\bm \zeta}^{(2)}(z,\tilde{x},\tilde{t})+\ldots
\,.
\end{array}
\label{eq:chexpans}
\end{equation}

\medskip

The next step to obtain a coarse-grained equation for the large-scale
dynamics is to plug \eqref{eq:chexpans} into
\eqref{eq:asym1}-\eqref{eq:asym3}. Exploiting the chain rule, the
definitions of $\xtil$ and $\ttil$ and averaging along $z$, we end up
with a set of equations involving solely the large-scale fields. The
equation for the large-scale transverse perturbation $\langle
w_z^{(1)} \rangle (\xtil,\ttil)$ is obtained from the solvability
condition at order $\epsilon^5$.  For details on the Newtonian case
and solvability conditions, see \cite{GVF94}.

We can summarize the whole procedure in the following schematic way:

\begin{enumerate} \item Solve the continuity equation. The explicit
expression of $w_z^{(n)}$ is thus obtained in terms of known functions
of $z$.

\item Solve the equation for $\zeta_{zz}^{(n)}$; this can always be
done algebraically as $\zeta_{zz}^{(n)}$ is slaved to the $w_z^{(n)}$
field.

\item Solve the evolution equation for $w_z^{(n)}$. This field is
obtained from ({\it a}); we are then able to obtain the expression for
the pressure field perturbation $q$.

\item Solve the system for $\zeta_{xz}^{(n)}$ and $w_x^{(n)}$ by
direct integration.

\item Algebraically obtain the explicit expression for
$\zeta_{xx}^{(n)}$.

\item Impose solvability condition at order $n+1$ on the continuity
and the velocity field equations. Such condition is automatically
fulfilled by the polymer conformation tensor, as it is slaved to the
velocity field at the previous order.  \end{enumerate}

\smallskip
The final equation has the form of a ``standard'' Cahn--Hilliard equation:
\begin{equation}\label{eq:Cahn-Hilliard}
\tilde{\partial}_t \langle w_z^{(1)} \rangle = \tilde{\partial}_x
\left[ \left(-A+B\langle w_z^{(1)}\rangle^2\right)\tilde{\partial}_x
\langle w_z^{(1)} \rangle\right] - C\tilde{\partial}_x^4\langle
w_z^{(1)} \rangle \quad .
\end{equation}

\begin{figure}
\centering
\includegraphics[angle=0, width=0.8\textwidth]{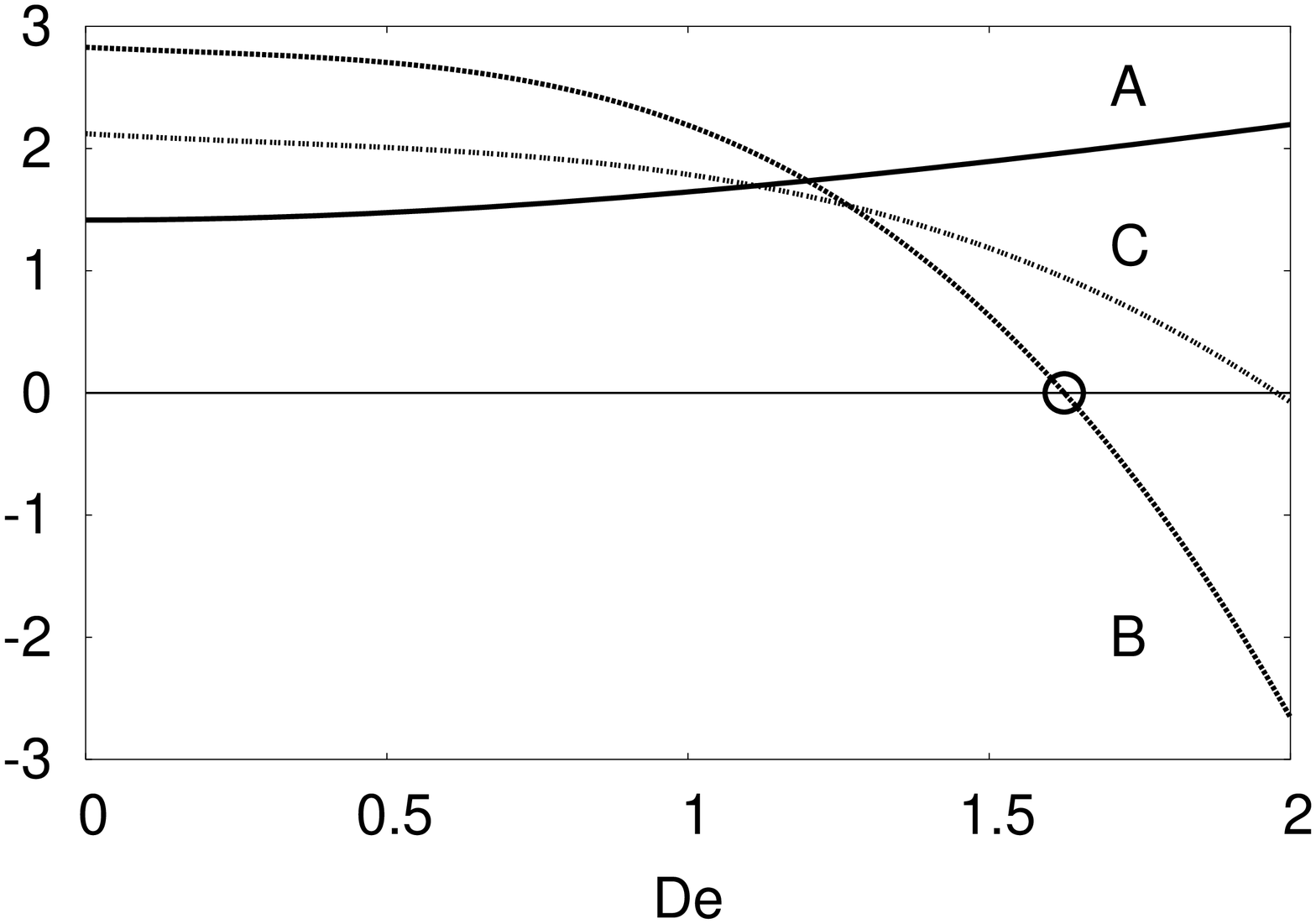}
\caption{The parameters $A$,$B$ and $C$ appearing in the
coarse-grained Cahn-Hilliard equation (\protect\ref{eq:Cahn-Hilliard}) as a
function of the Deborah number (for $\beta=0.769$). }
\label{fig:ABC}
\end{figure}

``Standard'' is meant to stress that the structure of
\eqref{eq:Cahn-Hilliard} (including the cubic nonlinearity) emerges in
a variety of hydrodynamic situations \cite[see
][]{N76,S85,P87,MY99}. The parameters $A,B,C$ are known functions of
the parameters $\De$ and $\beta$, as shown in figure~\ref{fig:ABC}. It
is worth noting that $A$ is non-negative as the system is supposed to
be slightly above the threshold of instability and we have explicitly
incorporated a negative sign in (\ref{eq:Cahn-Hilliard}).

The saturation of the instability requires two conditions. First, $C$
must be positive to ensure that the instability be saturated at
sufficiently high wave-numbers (still much smaller than those of the
basic flow, of order unity). Second, $B$ ought to be positive to
ensure that, as $\langle w_z^{(1)} \rangle$ becomes $O(\sqrt{A/B})$,
the nonlinear eddy-viscosity $-A+B\langle w_z^{(1)}\rangle^2$ change
sign and the growth be again saturated. Both these conditions are
satisfied up to a critical value of the Deborah number, $\De^*$ (see
figure~\ref{fig:ABC}).

Up to the critical Deborah number $\De^*$,
the equation (\ref{eq:Cahn-Hilliard}) for the large-scale
perturbations has the same structure as in the Newtonian case and the
only difference is in the numerical value of the parameters $A$, $B$,
$C$. As we shall see in the next sections, this property ceases to be
true above $\De^*$.

To conclude, we stress the fact that all the fields up to order four
are expressed in terms of explicit functions of the fast variables and
of the large-scale field $\langle w_z^{(1)} \rangle$, obeying the
Cahn--Hilliard equation (\ref{eq:Cahn-Hilliard}).


\section{Generalized Cahn--Hilliard dynamics}\label{sec:generalisedCH}

We have observed in the previous section that, along the marginal
linear stability curve, there exists a critical value of the Deborah
number, $\De^*$, where the cubic nonlinear term becomes
negative. Furthermore, the change of sign is taking place in the
region where the small-scale operator is stable. The problem is
thus well-posed and lends to multiscale methods.

For $\De > \De^*$, the field keeps growing at sufficiently large
scales, until it reaches amplitudes where the next-order nonlinearity
becomes important. Its structure is dictated by the conservation of
momentum and the symmetries of the basic flow: $\detil_x \left(\langle
w_z \rangle^4 \detil_x \langle w_z \rangle\right) $, with a regular
coefficient $D$ in the neighborhood of the critical point $P^*$,
where both the eddy-viscosity and the coefficient of the third-order
nonlinearity change sign.

\begin{figure} \centering 
\includegraphics[angle=0, width=0.8\textwidth]{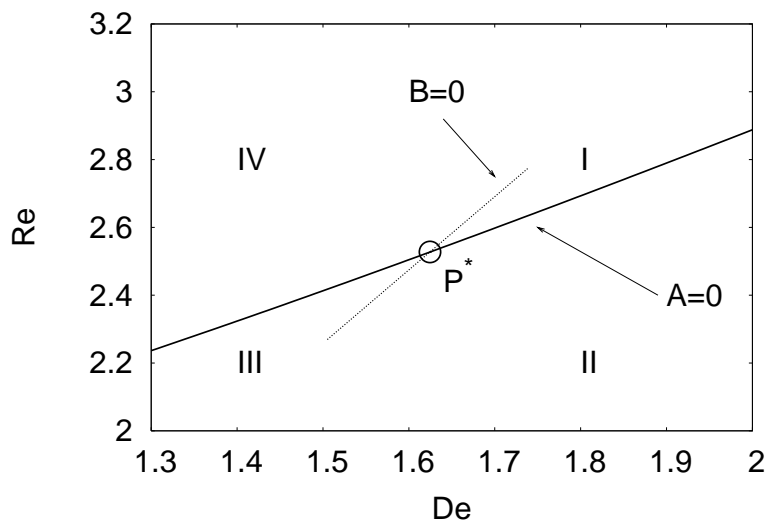} 
\caption{The
phase-space around the critical point $P^*$ where both the
eddy-viscosity and the coefficient of the third-order nonlinearity change
sign. The region is divided in four regions schematically sketched
here by the two critical curves $A=0$ and $B=0$. The former is found
from the linear stability analysis in section~3. The latter is found
locally, around the $A=0$ curve, by solving (\protect\ref{eq:Bexp}), and is
linearly extrapolated for graphical purposes as a dashed line. For
$\beta=0.769$, the curve $B=0$ is inclined at approximately $60^{\circ}$ with respect
to the $\De$ axis.}
\label{fig:zoneCHG} \end{figure}

Four regions can be identified around $P^*$ (see
figure~\ref{fig:zoneCHG}). The eddy-viscosity $A=0$ curve has been
obtained by means of the linear stability analysis
(\S\,\ref{sec:linearanalysis}). The linear approximation of the curve
$B=0$ in the vicinity of $P^*$ is obtained from the analytic
expression of $B$ on the marginal curve and the marginal curve itself.

Zone I is linearly unstable ($A>0$), has a third-order destabilizing
term ($B<0$) and we can guess that a fifth-order term will enter into
play to stabilize the growth. Zone II is particularly interesting as
it is linearly stable ($A<0$), but has a third-order destabilizing
contribution ($B<0$). Perturbing with a field of sufficiently strong
amplitude, the system jumps to the asymptotic steady state where the
two nonlinear terms (third and fifth-order) balance each other. Zone
III is completely stable ($A<0,B>0$). In the last region, IV, as $\De$
approaches the critical value, the coefficient $B$ goes to zero and
cannot saturate the exponential growth from the linear
instability. The fifth-order nonlinearity, which is negligible far
from the critical point, must enter again into play.

\subsection{Zone I}

When both the Reynolds and the Deborah numbers exceed their critical
values, previous considerations suggest the following structure for
the coarse-grained equation:
\begin{equation}\label{eq:Cahn-Hilliard-Gen} \tilde{\partial}_t w = -
A \tilde{\partial}^2_x w - |B| \tilde{\partial}_x (w^2 \tilde{\partial}_x
w) - C \tilde{\partial}_x^4 w + D \tilde{\partial}_x ( w^4
\tilde{\partial}_x w) \, .  \end{equation}

Confining the analysis to the surroundings of the critical point
$P^*$, we may represent the position in phase space as:
\begin{eqnarray}\label{eq:nuexp}
\nu=\nu^*(1-K_1\epsilon_{\nu}-K_2\epsilon_{\nu}^2)\,, \\ \label{eq:deexp}
\De=\De^*(1+\epsilon_{\tiny{\De}}) \, .  \end{eqnarray} Adequately choosing the
$K_1$ and $K_2$ parameters, any point around $P^*$ can be reached as
$\epsilon$ varies. The reason why we need to incorporate in
\eqref{eq:nuexp} the additional contribution of order $\epsilon^2$
will be clear shortly. 

In the neighborhood of $P^*$, the coefficients $A$ and $B$ are expanded as:
\begin{eqnarray}\label{eq:Aexp} A=\frac{\partial A}{\partial \De} (\De
-\De^* ) +\frac{\partial A}{\partial \nu} (\nu -\nu^*)\,,
\\\label{eq:Bexp} B=\frac{\partial B}{\partial \De} (\De -\De^* )
+\frac{\partial B}{\partial \nu} (\nu -\nu^*) \,,
\end{eqnarray}
where all derivatives are computed at $P^*$.

The scaling in $\epsilon$ of the velocity field amplitude,
$\epsilon_w$, and the parameters $\epsilon_{\nu}$, $\epsilon_{\tiny{\De}}$ is
found by requiring that all terms in \eqref{eq:Cahn-Hilliard-Gen} be
of the same order in the scale-separation small parameter $\epsilon$.

The comparison between the last two terms in
(\ref{eq:Cahn-Hilliard-Gen}) fixes the relation between $\epsilon$ and
$\epsilon_w$:
\begin{equation}\label{eq:epsepsw} D ~ \epsilon^2 \epsilon_w^5 \sim C
~ \epsilon^4 \epsilon_w \Rightarrow \epsilon_w = \epsilon^{1/2} \, .
\end{equation}

The parameters $\epsilon_{\nu}$ and $\epsilon_{\tiny{\De}}$ are found by
comparing the terms associated to $A$, $B$ and $D$ in
(\ref{eq:Cahn-Hilliard-Gen}). Using
\eqref{eq:nuexp}-\eqref{eq:epsepsw}, we obtain:
\begin{eqnarray}\label{eq:bilanciaAD} D ~ \epsilon^2 \epsilon^{5/2} \sim
\left[\frac{\partial A}{\partial \De} (\epsilon_{\tiny{\De}} \De^* ) -
\frac{\partial A}{\partial \nu} (K_1\epsilon_{\nu} \nu^*) \right]
\epsilon^2 \epsilon^{1/2} \\
\label{eq:bilanciaBD} D ~ \epsilon^2 \epsilon^{5/2} \sim \left[
\frac{\partial B}{\partial \De} (\epsilon_{\tiny{\De}} \De^* )
-\frac{\partial B}{\partial \nu} (K_1\epsilon_{\nu} \nu^*) \right]
  \epsilon^2 \epsilon^{3/2} \, .  \end{eqnarray} 

Choosing $\epsilon_{\nu}=\epsilon_{\tiny{\De}}=\epsilon$ and setting $K_1$ to
ensure $\left[ \frac{\partial A}{\partial \De} \De^* - \frac{\partial
 A}{\partial \nu} K_1 \nu^* \right] = 0$, 
both equations \eqref{eq:bilanciaAD} and \eqref{eq:bilanciaBD} are
satisfied. Equation \eqref{eq:bilanciaAD} is balanced by the
second-order term of the $\nu$ expansion (\ref{eq:nuexp}), dependent
on $K_2$. The scalings of time, pressure and polymer conformation tensor
perturbation, $\epsilon^4$, $\epsilon^{1/2}$ and $\epsilon^{1/2}$,
respectively, are derived as in \S\,\ref{sec:standardCH}.

\medskip
Once the scalings have been determined we can proceed as in
\S\,\ref{sec:standardCH} to obtain the large-scale equation for
$\langle w_z^{(1/2)} \rangle(\ttil,\xtil)$. The evolution equation
emerges now from the solvability condition at the order
$\epsilon^{9/2}$:
\begin{equation}\label{Cahn-Hilliard-Gen-SS} \tilde{\partial}_t
\langle w_z^{(1/2)} \rangle = \tilde{\partial}_x \left[
\left(-A+B\langle w_z^{(1/2)}\rangle^2 + D\langle w_z^{(1/2)}\rangle^4
\right)\tilde{\partial}_x \langle w_z^{(1/2)} \rangle\right] -
C\tilde{\partial}_x^4\langle w_z^{(1/2)} \rangle \, ,
\end{equation} where the coefficients are explicit functions of
$\beta$. For $\beta=0.769$, they read:
\begin{equation}\label{JJJ} \left \{ \begin{array}{ll} A=0.5106 +1.965K_2\,,\quad
B=-8.979\,, \\ C=0.9439\,, \quad D=23.11\,, \quad K_1=0.594\,.\end{array} \right \} 
\end{equation}

Although \eqref{Cahn-Hilliard-Gen-SS} belongs to the class of the
Cahn--Hilliard equations \eqref{landau}, the emergence of the new,
sixth-order nonlinearity will be responsible for new dynamical
aspects, not present for $\De < \De^*$, which will be discussed in
detail in \S\,\ref{sec:variationalf}.

\subsection{Zone II}

For Deborah numbers above the critical value, perturbations are
nonlinearly unstable: $B<0$.  This is true regardless of the sign of
the linear term and strong enough perturbations may then grow even if
the system is linearly stable.

Let us then consider systems with $\nu > \nu^*$ and $\De > \De^*$. No
major difference with respect to case I is expected. At zero-th order,
the coefficients $A$ and $B$ vanish and equations
\eqref{eq:Aexp}-\eqref{eq:Bexp} hold. Again, we define the position in
phase-space via the two parameters $\epsilon_{\nu}$ and
$\epsilon_{\De}$. As the viscosity is now larger than the critical
value, a positive sign appears in the expansion of the viscosity:
\begin{equation}\label{eq:nuexp2}
\nu=\nu^*(1+K_1\epsilon_{\nu}-K_2\epsilon_{\nu}^2)\,,
\end{equation} 
while (\ref{eq:deexp}) holds. The parameter $K_2$, as we shall point out
later, can take any value compatible with the condition $A>0$.

The same calculations discussed in the previous subsection can be
carried out to derive the coarse-grained equation for the transverse
velocity. As one might expect, its form is exactly the same as
\eqref{Cahn-Hilliard-Gen-SS}, a generalized Cahn--Hilliard
equation. The only difference is in the value of the parameters. For
$\beta=0.769$, they read:
\begin{equation}
\left \{
\begin{array}{ll}
A=-0.2202 +1.965K_2\,,\quad
B=-35.62\,, \\
C=0.9439\,, \quad
D=23.11\,, \quad
K_1=0.5974\,.
\end{array}
\right \}
\end{equation}

Only $A$ and $B$ have changed with respect to \eqref{JJJ}, as expected
since they are the only parameters which depend on $\epsilon$ (and
thus on $\Rey$ and $\De$) in physical coordinates. Notice that there
is an upper bound on the values we can choose for $K_2$, reflecting the
linear stability requirement.

\subsection{Zone IV}

What happens when the Deborah number is barely smaller than the
critical value $\De^*$? Sufficiently close to it, the third-order
instability can be made subdominant with respect to the fifth-order
and our aim here is to work out the scaling coefficients corresponding
to such situation.

For this purpose, let us assume that the cubic nonlinearity is
negligible. At leading order, the terms associated to $A$, $C$ and $D$
must be of the same order. This means:
\begin{eqnarray} \label{eq:bilanciaADIII}
\epsilon^4 \epsilon_w \sim \epsilon^2 \epsilon_w \left[ \frac{\partial
 A}{\partial \De} (\De- \De^* ) + \frac{\partial A}{\partial \nu}
 (\nu- \nu^*) \right]\,, \\ \label{bilanciaACIII} \epsilon^2
 \epsilon_w^{5} \sim \epsilon^2 \epsilon_w \left[ \frac{\partial
 A}{\partial \De} (\De- \De^* ) + \frac{\partial A}{\partial \nu}
 (\nu- \nu^*) \right] \, ,
\end{eqnarray}
and implies:
\begin{equation}
\nu=\nu^*(1-K_2\epsilon^2)\,,\qquad\De=\De^*(1-\epsilon^2) \nonumber \, .
\end{equation}
Additionally, the velocity field scales as $\epsilon^{1/2}$, as the
pressure and polymer fields do. The time derivative scales as
$\epsilon^4$.

To be consistent, we are left to check that the third-order
nonlinearity is negligible. Using the previous scalings and the
ensuing fact that $B\sim O(\epsilon^2)$, we have to verify that:
\begin{equation}
O(B \partial^2 w^3) \ll O (D \partial^2 w^5) \,\,\Rightarrow
O(\epsilon^{11/2}) \ll O(\epsilon^{9/2}) \nonumber \, ,
\end{equation}
which holds true. It is now possible to apply the strategy
discussed in \S\,\ref{sec:standardCH} to derive the large-scale
equation and obtain (at order $\epsilon^{5}$):
\begin{equation}\label{Cahn-Hilliard-Gen-zonaIV}
\tilde{\partial}_t \langle w_z^{(1/2)} \rangle = \tilde{\partial}_x
\left[ \left(-A + D\langle w_z^{(1/2)}\rangle^4
\right)\tilde{\partial}_x \langle w_z^{(1/2)} \rangle\right] -
C\tilde{\partial}_x^4\langle w_z^{(1/2)} \rangle \, ,
\end{equation}
where $C$ and $D$ have the same value as before and $A=1.1740+1.965 K_2$.

\section{Variational formulation}\label{sec:variationalf}

It is well-known that the ``standard'' Cahn--Hilliard equation admits
a variational formulation in terms of a Ginzburg--Landau potential
\cite[][]{CH58}. Equation~\eqref{eq:Cahn-Hilliard}, after appropriate
rescalings, $w \rightarrow (A/B)^{1/2} w$, $t \rightarrow A^{-1} t,
\lambda= C/A$, is recast in the form \eqref{landau} with the Lyapunov
functional:
\begin{equation}
F[ w ] = \int \left[ \frac{\lambda}{2} ( \partial_x w )^2 +
I(w)\right] {\rm d}x \,, \qquad I(w)= -\frac{w^2}{2} +\frac{w^4}{12}
\, .
\end{equation}
Note that mean fields only are considered, that is $w$ must be read as
the rescaled leading contribution $\langle
w_z^{(1)}\rangle(\xtil,\ttil)$.

The existence of a Lyapunov functional implies the existence of an
asymptotic state for $w$, if boundary conditions are periodic and
stationary. Such state corresponds to a minimum of the Lyapunov
functional and it is calculated by the following equations:
\begin{equation}\label{eq:kinksh}
I'(w)= \lambda \partial_x^2 w \leftrightarrow \partial_x I=
\frac{\lambda}{2} \partial_x ( \partial_x w)^2\, .
\end{equation}

When $w$ is a maximum (or a minumum),
$I(w_{max})$ is constant and $w_{max}$ is obtained solving
$I'=0$ as $w_{max}=\pm \sqrt{3}$. Considering this boundary
condition, equation \eqref{eq:kinksh} can be easily solved. Its solutions
are the well-known \emph{kink} and \emph{anti-kink} structures,
namely:
\begin{equation} \label{kink}
w= \pm \sqrt{3} \tanh \left[ \sqrt{\frac{1}{2 \lambda}} x \right] \, .
\end{equation}

\medskip

The issue now is whether or not a Lyapunov extremal formulation exists
in the generalized Cahn--Hilliard case (\ref{eq:Cahn-Hilliard-Gen}) as
well, and how it relates to the standard one. In particular, a
Painlev\'e test \cite[][]{AC91} can be performed on the equation to
check its integrability.  The calculation consists in checking that
all movable singularities (whose location depends on initial and/or
boundary conditions) are poles \cite[see for details][]{AC91}. The
test is based on a well-known connection between the integrability
property of a nonlinear differential equation and its analytic
structure for complex values of the independent variable
\cite[][]{K889,K890,P897}. The explicit calculation is performed in
Appendix \ref{app:painleve}. The generalized Cahn--Hilliard equation
indeed enjoys the Painlev\'e property and is thus integrable.

Let us then write the equation (\ref{eq:Cahn-Hilliard-Gen}) after the
rescalings $w \rightarrow (A/B)^{1/2} w$, $t \rightarrow A^{-1} t,
\lambda= C/A, \gamma=A\,D/B^2$:
\begin{equation}\label{eq:gchresc}
\partial_t w = -\partial_x^2 w -\frac{1}{3}\partial_x^2 w^3 - \lambda
\partial_x^4 w + \frac{\gamma}{5} \partial_x^2 w^5 \, .
\end{equation}
Integrability of this equation is related to the existence of the
following Lyapunov functional, similar to that of the standard case,
yet with a sixth-order potential:
\begin{equation}
\label{eq:CHGpot}
F[w] = \int \left[\frac{\lambda}{2} ( \partial_x w )^2 + I(w)\right]
{\rm d}x\,, \qquad I(w)= -\frac{w^2}{2} -\frac{w^4}{12} +
\frac{\gamma}{30}w^6 \, .
\end{equation}
The calculation of the function corresponding to its minimum is
performed using again \eqref{eq:kinksh}.

\begin{figure}
\centering
\includegraphics[angle=0,width=0.8\textwidth]{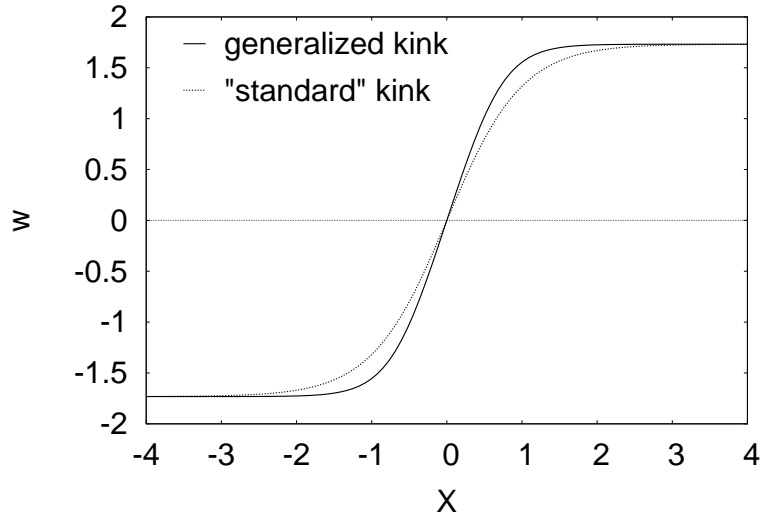}
\caption{The ``generalized'' (solid) and ``standard'' (dotted) kinks
for $\lambda=1/2$ and $\gamma=10/9$. The former has a manifestly
shorter range. It is shown in the body of the text that this entails 
longer time-scales for their annihilation with the corresponding 
anti-kinks.}
\label{fig:kinks}
\end{figure}

All solutions tend to final steady states which minimize $F$. The
approach to the solution is however nontrivial and the structure is
made of plateaux having velocity $\pm W_0$ ($I'(W_0)=0$), separated by
positive and negative kinks (see figure~\ref{fig:kinks}). The
amplitude of the velocity $w$ in the plateaux is:
\begin{equation} \label{eq:U2}
W_0^2 = \frac{  5+\sqrt{25+180 \gamma}   }{ 6 \gamma} \, .
\end{equation}
Note that, at small $\gamma$'s, the asymptotic velocity $W_0$ diverges
as $1/\sqrt{\gamma}$. This is quite intuitive: the field amplitude
equilibrating the third and the fifth-order nonlinearities increases
as the coefficient of the fifth-order nonlinearity reduces.  

The explicit expression of the profiles for kinks and anti-kinks is
obtained from the integration of equations
\eqref{eq:kinksh},\eqref{eq:CHGpot}. For the sake of example, when
$\lambda=1/2$ and $\gamma=10/9$ the profiles read:
\begin{equation} \label{kinkGEN}
w= \pm \sqrt{15} \frac{e^{2 \sqrt{3} x}-1 } { \sqrt{ 5 e^{4 \sqrt{3} x
} + 26 e^{2 \sqrt{3} x} + 5 } } \, .
\end{equation}

The generalized kink-antikink structures, e.g. those given by equation
\eqref{kinkGEN}, will be dubbed ``generalized'' kinks and anti-kinks.

\begin{figure}
\centering
\includegraphics[angle=0,width=0.8\textwidth]{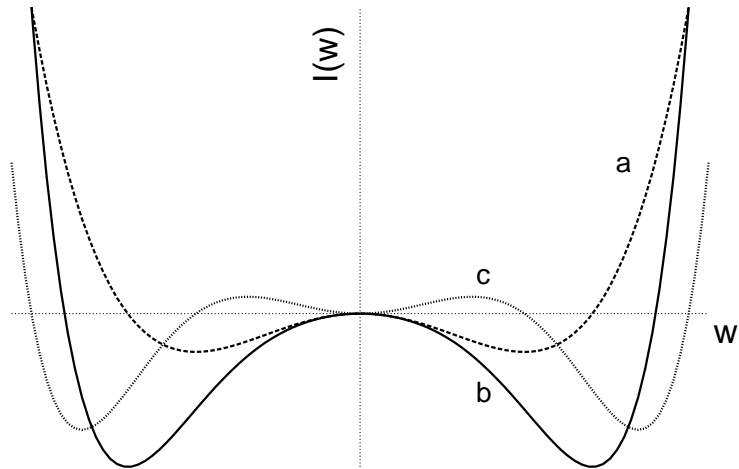}
\caption{The potentials associated to the different evolution
equation. Curve $a$ is related to the standard Cahn--Hilliard equation
(fourth-order potential); curve $b$ represents the generalized
Cahn--Hilliard equation (sixth-order potential). Curve $c$ is the
characteristic triple-well potential of the purely nonlinearly
unstable case. The plots are in arbitrary units, to ease the
comparison between the curves.}
\label{fig:potentials}
\end{figure}

\subsection{Dynamics of generalized kink/antikink annihilation and 
approach to equilibrium}

Detailed calculations are performed following \cite{LV03}, who in
turn based theirs on \cite{KO82}. They are lengthy, yet quite simple in
their basic idea, sketched hereafter. 

During metastable transitions,
the kinks do not satisfy \eqref{eq:kinksh} exactly, due to the
presence of other kinks and/or anti-kinks. The deviation of the
amplitude in the plateau is proportional to $e^{-s\Lambda}$, where
$\Lambda=4|x|$ and $x$ denotes the distance to the point where
$w=0$. Here, $s$ is the inverse of the typical length scale of this
deviation \cite[for details, see Appendix A of][]{LV03}. The quantity
$s$ turns out to be crucial as neighboring kinks and anti-kinks attract
proportionally to $e^{-s \Delta x}$, where $\Delta x$ is the distance
between neighbouring kinks and anti-kinks \cite[for details, see
Appendix B of][]{LV03}.

The behavior of the kink size $s$ is grasped as
follows. Consider a metastable state of the Cahn--Hilliard
equation. The potential felt by a kink $w(x)$ close to the plateau
$w=W_0$ is estimated by the Taylor expansion:
\begin{equation}\label{potexp} I(w-W_0) \simeq I(W_0) + I'(W_0) (w-W_0) +
I''(W_0) \frac{(w-W_0)^2}{2} \, , \end{equation} where we know that
$I'(W_0)=0$. Note also that the dynamics of $w$ does not change if
we add an arbitrary constant to the potential $I$, so that
we can set $I(W_0)\equiv 0$. 

Let us now calculate the shape of the profile
between $w$ and $W_0$. For a metastable state, $\partial_t (w-W_0)=0$,
that implies: \begin{equation} \frac{\lambda}{2}( \partial_x (w-W_0) )^2 +
[ I''(W_0)\frac{(w-W_0)^2}{2} ] = 0 \, .
\end{equation}
Interpreting $\partial_x$ as the inverse of the typical length scale
$s$ for $(w-W_0)$, one easily obtains $s=\sqrt{\lambda/
I''(W_0)}$. The second-order derivative can be explicitly calculated
using \eqref{eq:U2}:
\begin{equation}\label{eq:I2chg}
I''(W_0)=4+{\frac{2}{3}}W_0^2 \, .
\end{equation}

Qualitative properties of $s$ are easy to grasp.  At large $\gamma$'s,
the size of the kinks tends to a constant, independent of $\gamma$.
At small $\gamma$'s, the kinks get steeper and steeper, their size
scaling as $\gamma^{1/2}$. This implies that the convergence to
equilibrium will be slower and slower as $\gamma$ is reduced (recall
that the kinks attract proportionally to $e^{-s \Delta x}$).

\medskip
For the same band of unstable modes, \ie~keeping $\lambda$ fixed, it
holds that the convergence to equilibrium is slower for the
generalized than for the standard Cahn--Hilliard equation. Indeed, for
the Cahn--Hilliard potential $I_{CH}=-{w^2 / 2} + {w^4 / 12}$, the
second-order derivative $I_{CH}''(W_0)=2$. As for \eqref{eq:I2chg}, we
can use the identity $1+W_0^2/3=\gamma W_0^4/5$, following from the
very definition $I'(W_0)=0$, to obtain $I''(W_0)>2$.  This implies
that the interactions for the generalized kink-antikink structures
have a shorter range and their dynamics of annihilation is thus
slower.

A special remark applies to the linearly stable case (zone II). In
this case, the equation is associated to an uncommon triple-well
potential. The typical nonlinear kink-antikink dynamics appears only
if the initial perturbation will be energetic enough to let the system
``jump'' out of the central well and fall into one of the side wells.

\section{Numerical results}\label{sec:numres}

The analytical results presented in this work have been obtained by
multiscale techniques. Their basic assumption is the strong scale
separation between the basic flow and the most unstable perturbations.
In this section, we shall present the numerical simulations
performed to check the validity of this assumption. 

The linear analysis results have already been checked in \cite{B05} by
reducing the original linear partial differential equation to a
generalized eigenvalue problem and computing its
eigenvalues/eigenvectors. The scale separation hypothesis is found to
be well verified up to Deborah numbers of order unity ($\De \approx
2.3$ for the value $\beta=0.769$ used in this study). 
For larger $\De$, scale separation does not
hold and multiscale methods are not applicable.

To check the nonlinear results derived here, we have numerically
simulated the complete Oldroyd-B model equations
\eqref{eq:Diffuu}-\eqref{eq:Diffusigma} via a pseudospectral algorithm
\cite[see][for details on the numerical method]{CHQZ88}. In the following, we
will refer to these as to Direct Numerical Simulations (DNS), while numerical
integration of the one-dimensional Cahn--Hilliard equation
\eqref{eq:Cahn-Hilliard} will be referred to as CH simulations.  

To enforce a transverse perturbation, we integrated the equations on a
rectangular slab with $L=L_x=2 \pi$ and $L_z=64 L_x$. The aspect ratio
$r$ is then fixed at $1/64$. The simplest check of our results
concerns the growth rates of the instability which, in the linear
regime, can be obtained by the Cahn--Hilliard equation.  Neglecting
the nonlinear term, the dispersion relation for the transverse Fourier
modes $k$ reads:
\begin{equation}\label{eq:growthrate} 
g = A \left(\frac{\Rey}{\Rey_c}-1\right) k^2 - C k^4 \end{equation}

\begin{figure}
\centering 
\includegraphics[angle=0, width=0.8\textwidth]{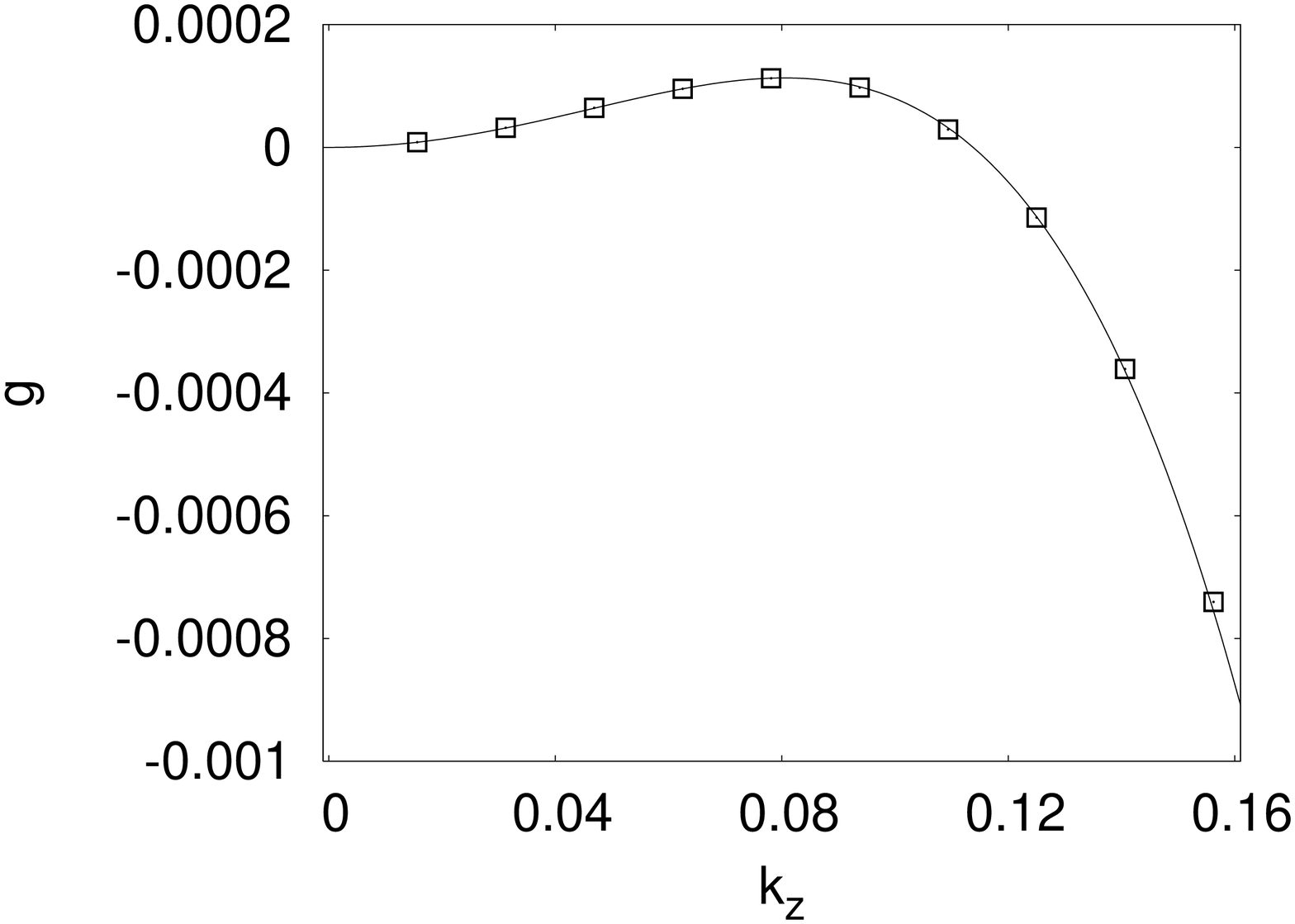}
\caption{Growth rates $g$ of the transverse Fourier modes $k$ for a
simulation with $\De=1.4$ and $\beta=0.769$. The simulations are
performed in a rectangular domain with aspect ratio $1/64$. The distance
to the critical point $\Rey/\Rey_c-1$ is $0.28$. The solid line represents
the linear prediction \protect\eqref{eq:growthrate}. The circles representing
the numerically computed growth rates have been
obtained with a DNS simulation by a linear fit of the logarithm of the energy for each mode
versus time, in the early stages of their exponential growth. }
\label{fig:growthrate}
\end{figure}

In figure~\ref{fig:growthrate}, we report the growth rates of the
first modes for a (white-noise in space) small initial perturbation.
We are then able to observe also negative $g$ (stable modes). The
comparison with the linear prediction is excellent, even for modes
whose scale separation is not very small.

Let us now consider the nonlinear stage of the perturbation growth. It
is well known that the time evolution of the Cahn--Hilliard equation
shows a succession of long-lasting metastable states characterized by
a well defined periodicity. For sufficiently small initial
perturbations, the wave-number $k$ associated to the maximum
growth-rate $g$ will be the first to reach the balance between the
destabilizing linear term $A\tilde{\partial}^2 \langle w_z \rangle$
and the stabilizing non-linear one $B\tilde{\partial}^2 \langle w_z
\rangle^3$.  When such equilibrium is reached, the energy associated
to that mode is constant and the system is quasi-stable.  In the
meanwhile the other modes $k_{max}-1, k_{max}-2,\ldots$ keep
growing. When the mode $k_{max}-1$ balances the two terms, the energy
associated to the mode $k_{max}$ drops. This new state is again
quasi-stationary and has a well-defined periodicity $k_{max}-1$. 

\begin{figure}
\centering
\includegraphics[angle = 0, width=0.8\textwidth]{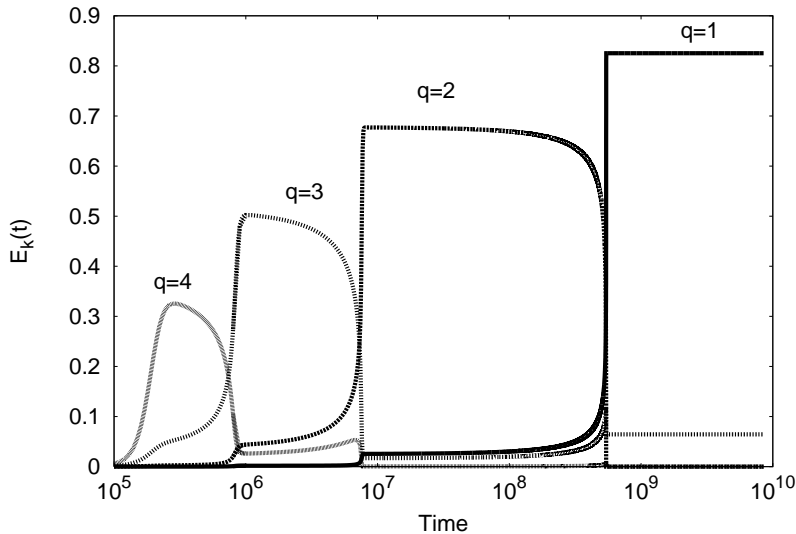}
\caption{The energy associated to the lowest wavenumber modes resulting from a
  CH simulation.
The values of the coefficients have been arbitrarily chosen for conveniency of
 display. The quasi-stationary states can be
clearly seen up to the asymptotic one corresponding to the largest
periodicity. In this simulation, $\De=1.4$ and $\Rey/\Rey_c-1=0.28$.}
\label{fig:CHdyn}
\end{figure}

\begin{figure}
\centering
\includegraphics[angle =0, width=0.3\textwidth]{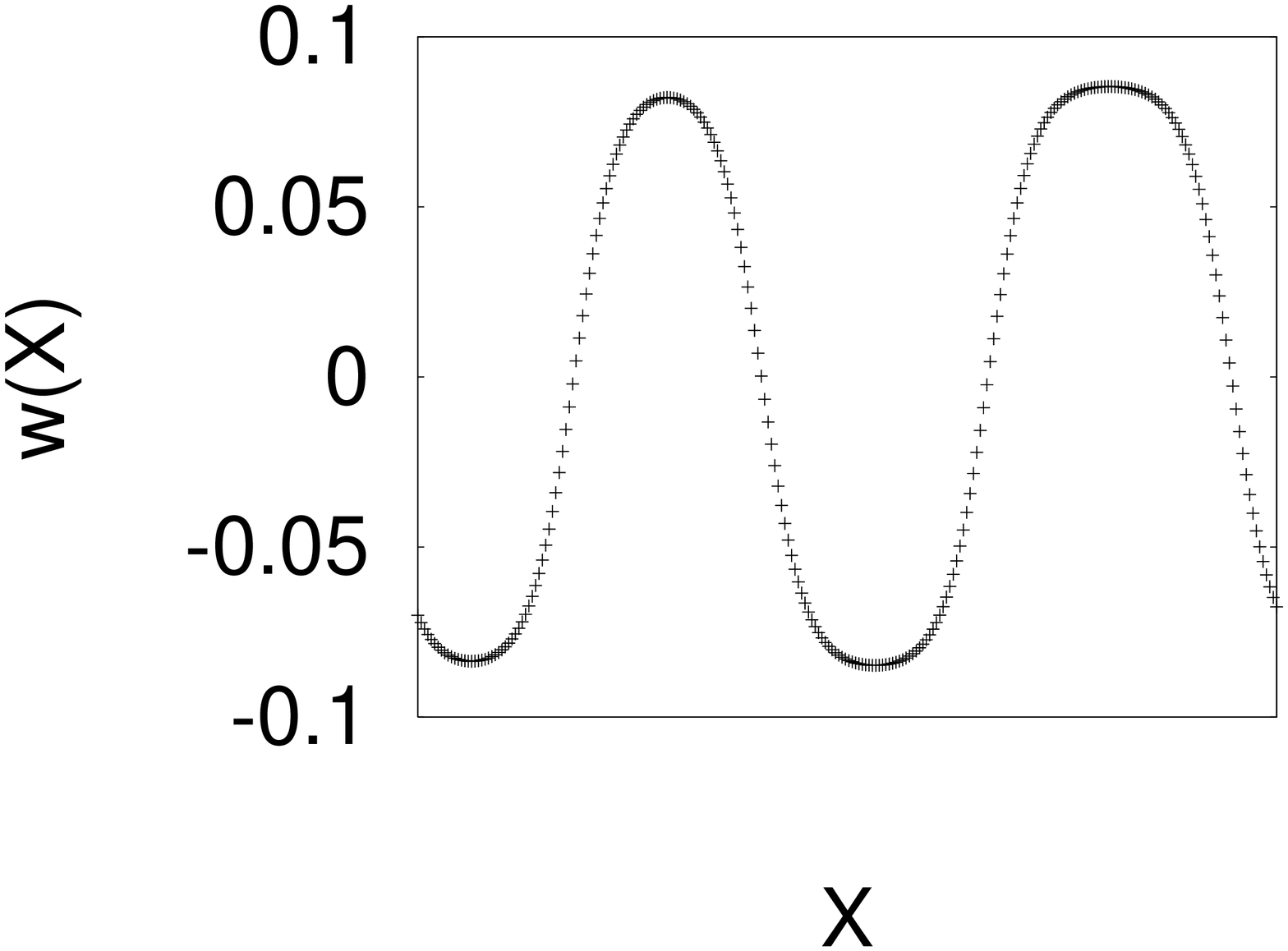}
\includegraphics[angle =0, width=0.3\textwidth]{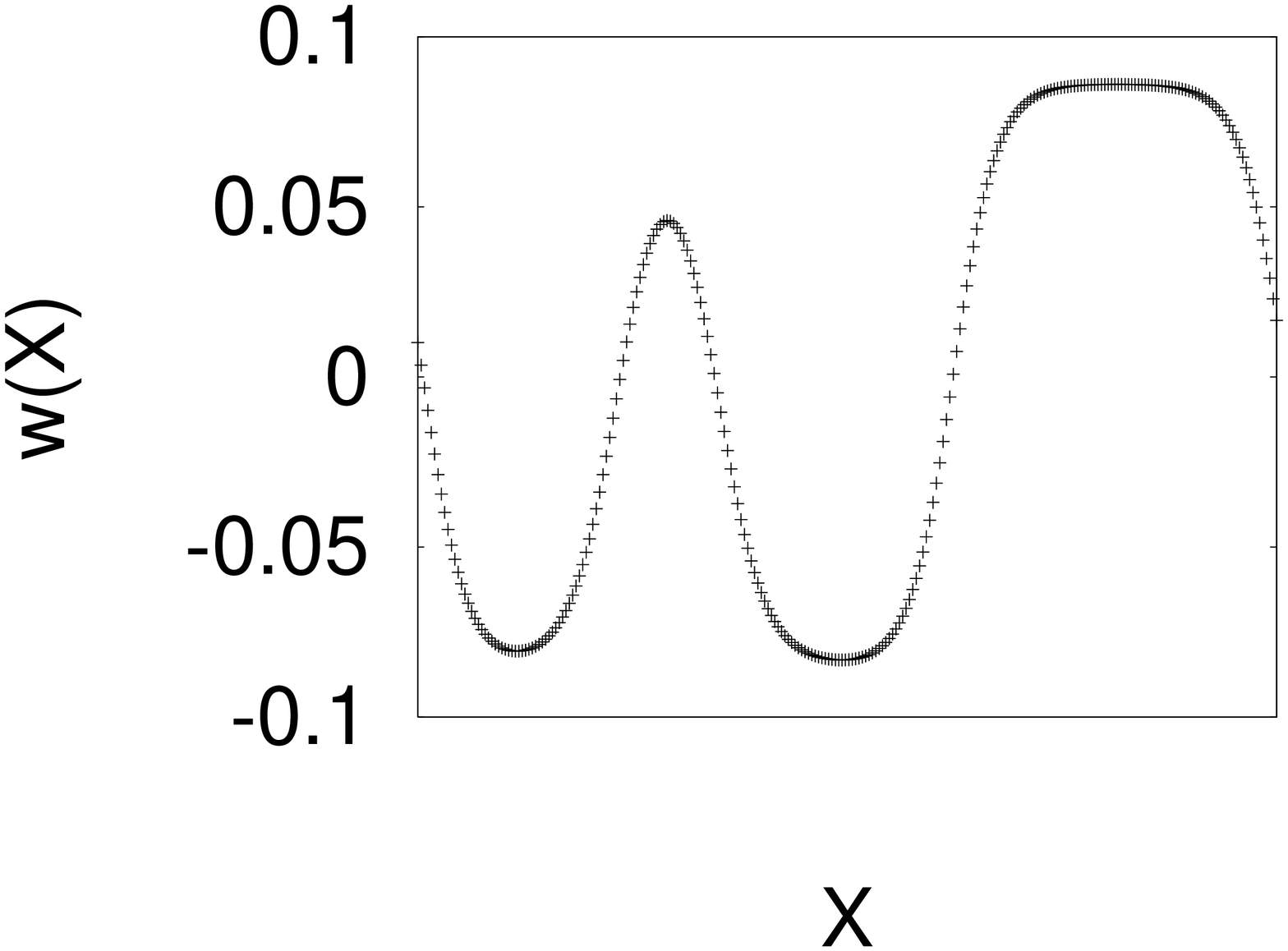}
\includegraphics[angle =0, width=0.3\textwidth]{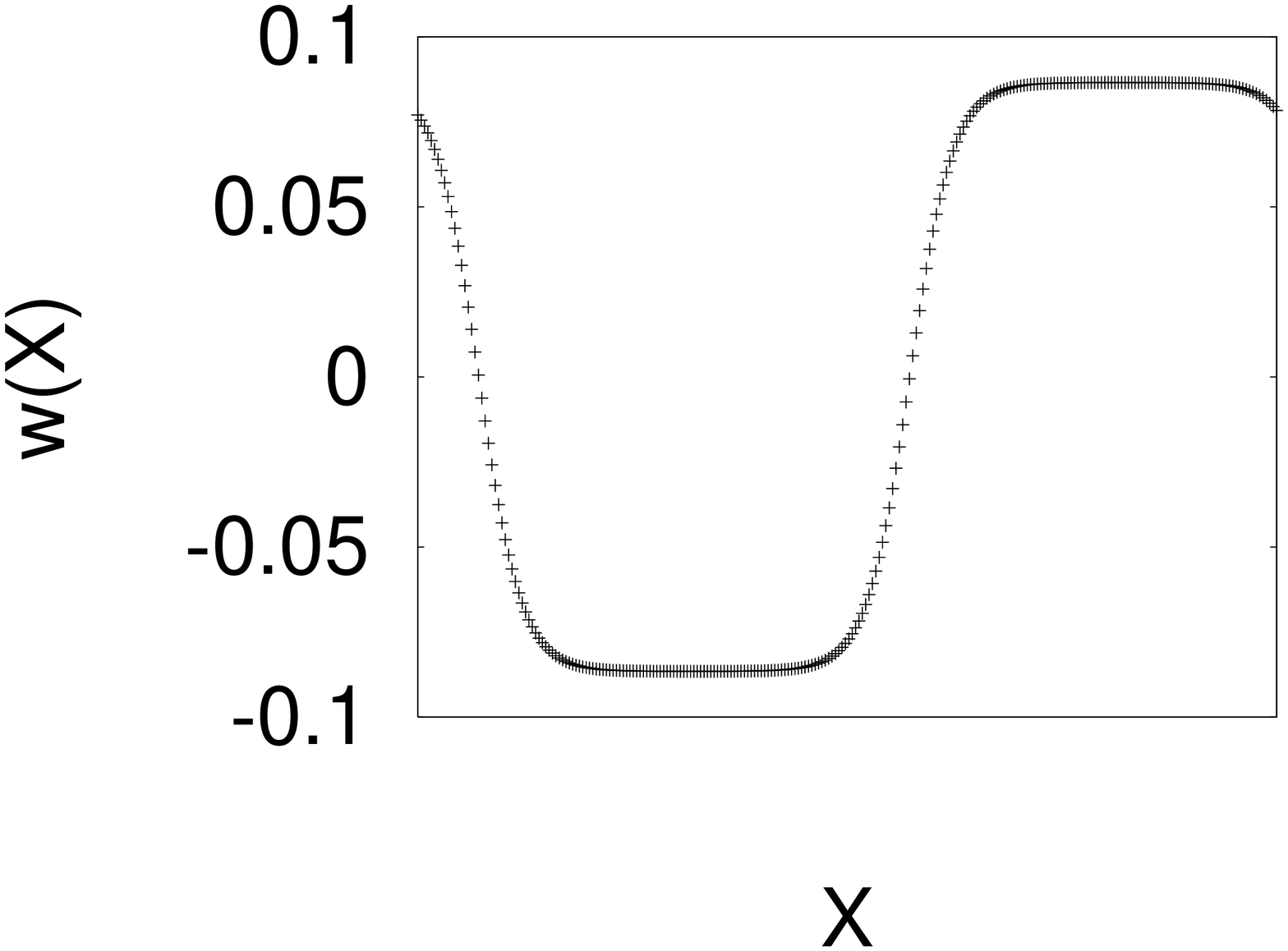}
\caption{Instantaneous transverse velocity field at different
times. The simulation is the same as in
figure~\ref{fig:CHdyn}. The transition between two metastable states
can be regarded as a kink-antikink annihilation. In this figure a
transition from a $k=2$ to a $k=1$ state is represented, the $x$-axis
being the physical $x$ direction of the integration box and the
$y$-axis being the amplitude of the $w$ perturbation. The time figure
set over the graphs refers to the evolution shown in
figure~\ref{fig:CHdyn}.} \label{fig:annihil}
\end{figure}

The process continues until a state with the box periodicity is reached
(see figure~\ref{fig:CHdyn}); such a state is stationary and
corresponds to the asymptotic behaviour in
\S\,\ref{sec:variationalf}. The kink structures described there, are
characteristic of all of these stages. Indeed, any transition between
two quasi-stationary states can be seen as a kink-antikink
annihilation, yielding a decrease in periodicity, as in
figure~\ref{fig:annihil} \cite[][]{She87}.

\begin{figure}
\centering
\includegraphics[angle=0, width=0.8\textwidth]{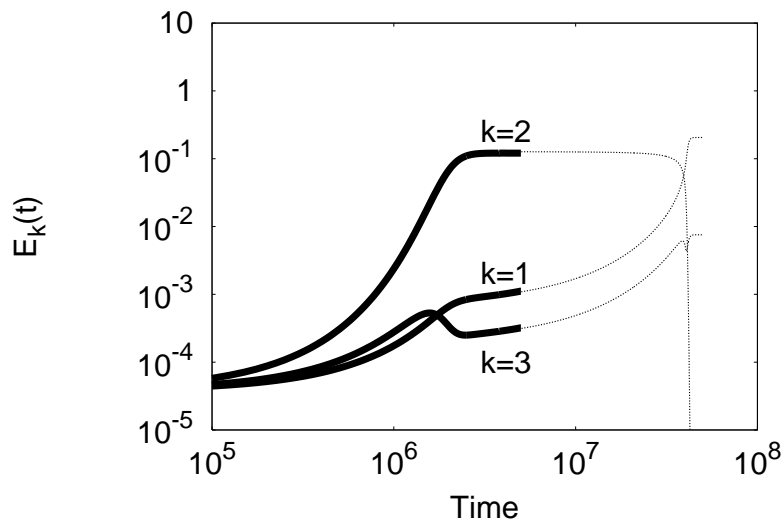}
\caption{The comparison between DNS simulations of the Oldroyd--B model
and the coarse-grained Cahn--Hilliard equation derived in the body of
the text. The thicker lines represent the evolution of the lowest-energy modes
in a DNS simulation, while the thinner lines are the result of a
CH simulation. Its dynamical parameters have been set with
the results obtained in section~\protect\ref{sec:standardCH}.
This particular figure refers to a simulation with
$\De=1.4$, $\beta=0.769$ and $\Rey/\Rey_c-1=0.14$.}
\label{fig:CHeDNS}
\end{figure}

To check the results obtained in \S\,\ref{sec:standardCH}, we have
performed a DNS simulation for a particularly long lapse of time.  The
excellent agreement between the DNS and the prediction of the
Cahn-Hilliard equation is shown in figure~\ref{fig:CHeDNS}.

\begin{figure}
\centering
\includegraphics[angle=0, width=0.8\textwidth]{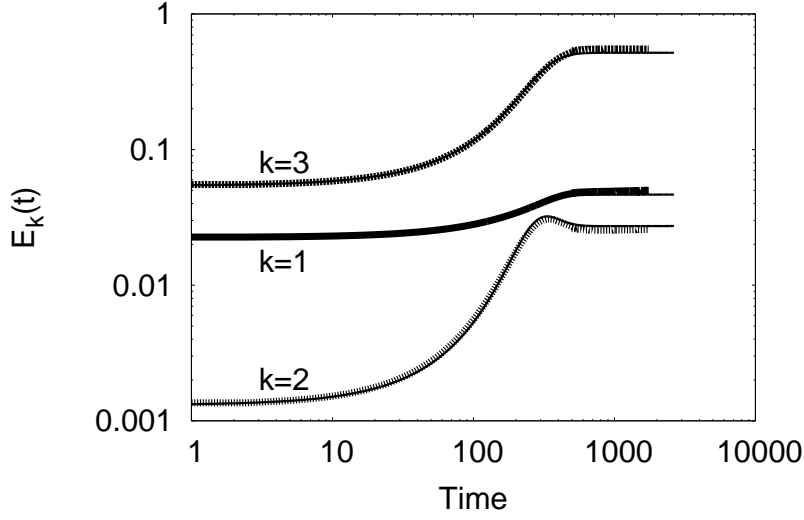}
\caption{The generalized Cahn--Hilliard equation reproduces the
dynamics of the Oldroyd-B model around the critical point $P^*$. As in
figure~\protect\ref{fig:CHeDNS}, the thicker lines are DNS simulations while the
thinner ones are CH simulations. This
comparison was realized for $\De=1.62$ and $\Rey=2.5159$.}
\label{fig:CHGENeDNS}
\end{figure}

The same comparison can be realized in the neighborhood of the
critical point $P^*$. This kind of simulation is much harder than for
the standard Cahn--Hilliard, because it involves a very precise
knowledge of the position of the critical point, and there is no easy
way to obtain this from the simulations. Moreover, any system we
simulate will be at a finite distance from the critical point. The
parameter that will mostly feel this difference will be $D$, as we
have chosen it to be approximately constant around $P^*$. We have been
able to overcome this weakness via a limited tweaking of the $D$
parameter in the CH simulation. As shown in
figure~\ref{fig:CHGENeDNS}, an excellent agreement between the curves
is again achieved.

\section{Clues on drag reduction}\label{sec:dragred}

One of the most striking properties of viscoleastic fluids is the drag
reduction effect. In 1949, Toms found that the injection of minute
amounts of polymers in turbulent fluids flowing in a channel was able
to increase the mean flow by an amount soaring up to 80\%. Even if
this phenomenon has been known for over fifty years
\cite[][]{T49,L69,V75}, a satisfactory understanding of its
fundamental mechanisms is still lacking.

A large number of experiments has been performed to study this effect
\cite[see, \eg][]{V75,NH95,SW00}, but a burst in its theoretical
analysis occured after drag reduction was found in numerical
simulations of viscoelastic fluids \cite[][]{SBH97}. The activity is
being spurred both by fundamental interest and industrial
applications \cite[][]{L92}.

Drag reduction is commonly associated to channel flows and boundary
effects.  Still, it is now clear that the phenomenon is present even
for free flows \cite[][]{BCM05}. What we show here is that, even at
relatively small Reynolds numbers, an increase in the Deborah number
produces an enhancement in the mean flow amplitude.  Simply looking at
the linear stability diagram (\ref{fig:stab}) we may already
conclude that, as the polymer elasticity grows, so does the critical
Reynolds number and the flow is stabilized. Let us further investigate
this effect analytically using the results of \S\,\ref{sec:standardCH}.

A parameter that can be employed to study the mean flow properties in
free flows is the drag coefficient $f$ \cite[][]{BCM05}:
\begin{equation} f=\frac{F_0 L}{U^2} \, .  \end{equation} The drag
coefficient can be seen as the ratio between the energy input (through
the forcing $F_0$) and the mean energy of the flow. As we are
interested in mean effects only, we will average $U^2$ over the basic
flow periodicity. This will ensure that only mean effects will be
taken into account.

When the state is linearly stable (low Reynolds numbers) we know that
no perturbation can alter the basic flow, $U=V=F_0 L^2/\nu$ and thus
$f=\Rey^{-1}$.

In \S\,\ref{sec:standardCH}, we have solved all the equations of motion
up to the fourth order. They give the following form of the flow (up to
the second order):
 \begin{eqnarray} \label{soluzWX}
 U_x(z) = V \cos(\frac{z}{L}) + \frac{V (L^2+(\beta-1) \nu \tau)}{\nu
 L} \langle w_z^{(1)}\rangle \sin{(\frac{z}{L})} + \\ - \frac{ \De [
 L^4 + \nu \tau (\beta-1) (2L^2 + \nu \tau \beta) ]}{\nu^2 \tau L}
 \langle w_z^{(1)}\rangle ^{2}\cos({\frac {z}{L}}) + \nonumber \\
 +\frac{ \De L (\beta-1)}{2} ( \detil_x \langle w_z^{(1)} \rangle )
 \sin (\frac {2z}{L} ) \, .\nonumber \end{eqnarray} The first term is
 the basic, stationary Kolmogorov flow. Averaging over all possible
 initial conditions, component proportional to $\sin(z/L)$ and
 $\sin(2z/L)$ disappear. The resulting expression for the mean flow
 in the $x$ direction reads:
 \begin{equation} \overline{U_{x}(z)} =(V + h(\De,\beta) \frac{
 \overline{\langle w_z^{(1)}\rangle^2} }{V} )\cos(\frac{z}{L})=
 V_{eff}\cos(\frac{z}{L}) \, , \end{equation} where the quantity
 $\overline{\langle w_z^{(1)}\rangle^2}$ follows from the
 Cahn--Hilliard equation in the stationary state:
\begin{equation}\label{CHsempl} 0 = - {\partial}^2_x \overline{\langle
w_z^{(1)} \rangle} A\epsilon^2 +\frac{B}{3}{\partial}^2_x
\overline{\langle w_z^{(1)}\rangle^3} - {\partial}^4_x
\overline{\langle w_z^{(1)} \rangle} C \, .  \end{equation} As
$\overline{\langle w_z^{(1)} \rangle}$ is periodic, we can integrate
twice over the domain and notice that, on the plateau, the last term is
zero. The field amplitude must then satisfy:
\begin{equation}\label{CHampl} 0 = - A\epsilon^2 +\frac{B}{3}
\overline{\langle w_z^{(1)}\rangle^2} \Rightarrow \overline{\langle
w_z^{(1)}\rangle}=\sqrt{\frac{3\epsilon^2A}{B}} \end{equation} Since
the analytical expression of $A$ and $B$ is known, as well as how
$\epsilon$ changes with $\De$ for a fixed Reynolds number, the 
analytical expression for $f$ is obtained: \begin{equation}
f=\frac{\nu F L}{ V_{eff}^2}=\frac{V^2}{\Rey V_{eff}^2} =
\frac{1}{\Rey (1+ h \frac{3A}{BV^2} \frac{\Rey-\Rey_c}{\Rey_c} )^2} \,
, \end{equation} where $h$, $A$, $B$ and $\Rey_c$ are explicit
functions of the Deborah number and $\beta$.

\begin{figure}
\centering
\includegraphics[angle=0, width=0.8\textwidth]{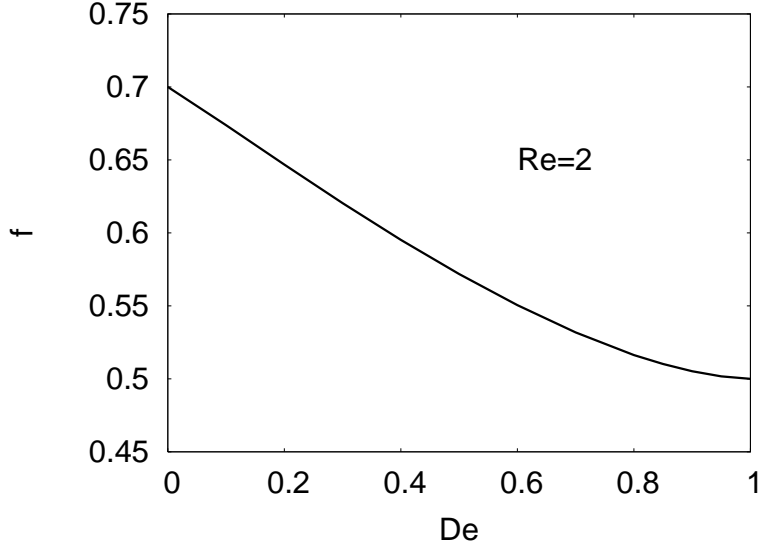}
\caption{The drag coefficient $f$ plotted versus the Deborah number
$\De$ at constant $\Rey$. As the polymer elasticity grows, the drag
coefficient diminishes. This implies that the mean flow grows with
$\De$. }
\label{fig:DR_TEO}
\end{figure}

As we want to investigate how the polymer elasticity affects the flow,
a meaningful approach is to keep the Reynolds number fixed, while
varying the Deborah number. This allows studying how the same flow
reacts when different kinds of polymers are injected. Once $\beta$ and
$\Rey$ are chosen, it is possible to plot $f$ versus $\De$ on the
basis of analytical results, as in figure~\ref{fig:DR_TEO}.
The drag coefficient is clearly decreasing with the Deborah number
even though the flow is barely in its nonlinear regime.

\begin{figure} \centering 
\includegraphics[angle =0, width =0.45\textwidth]{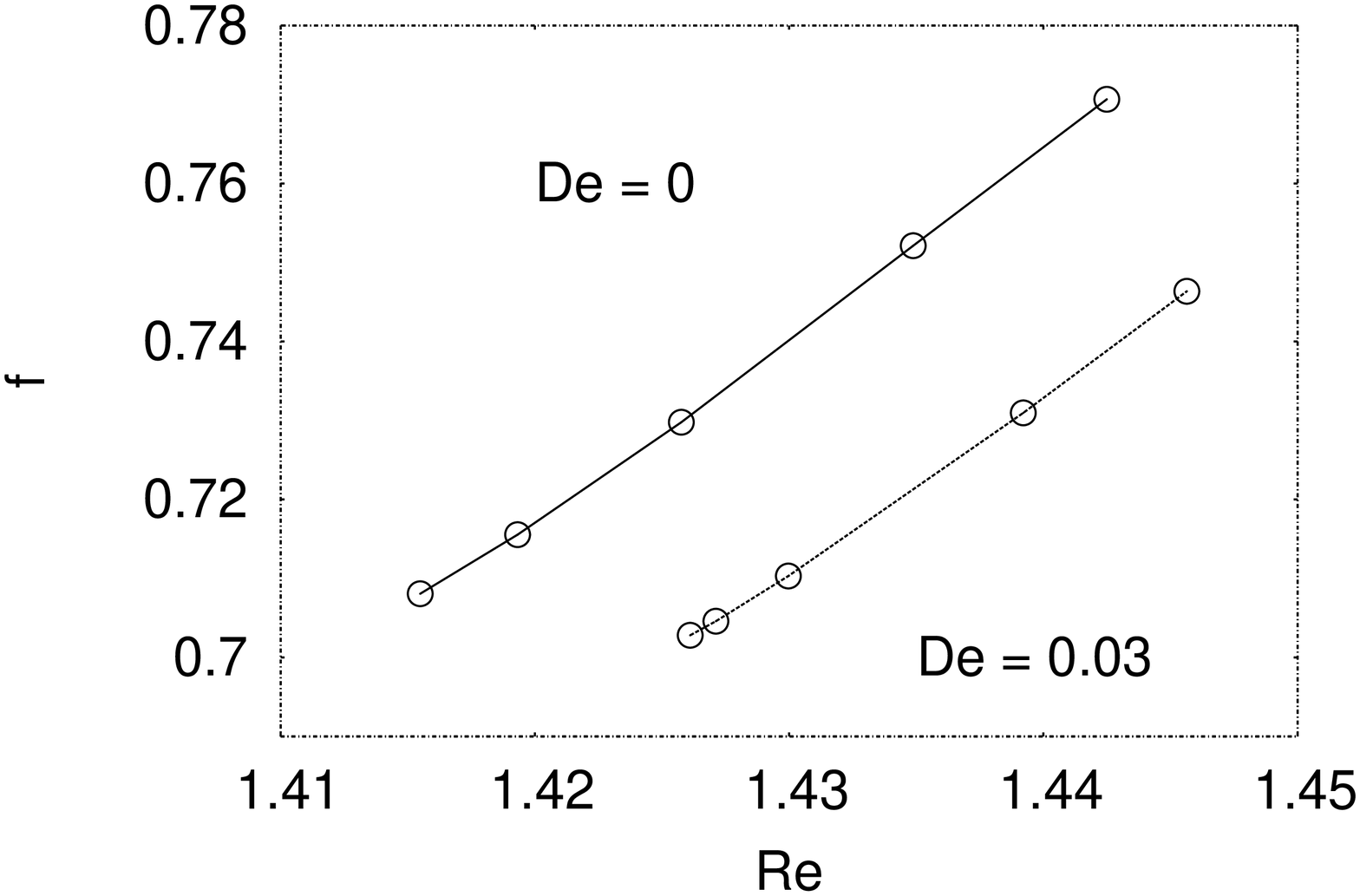} 
\includegraphics[angle =0, width =0.45\textwidth]{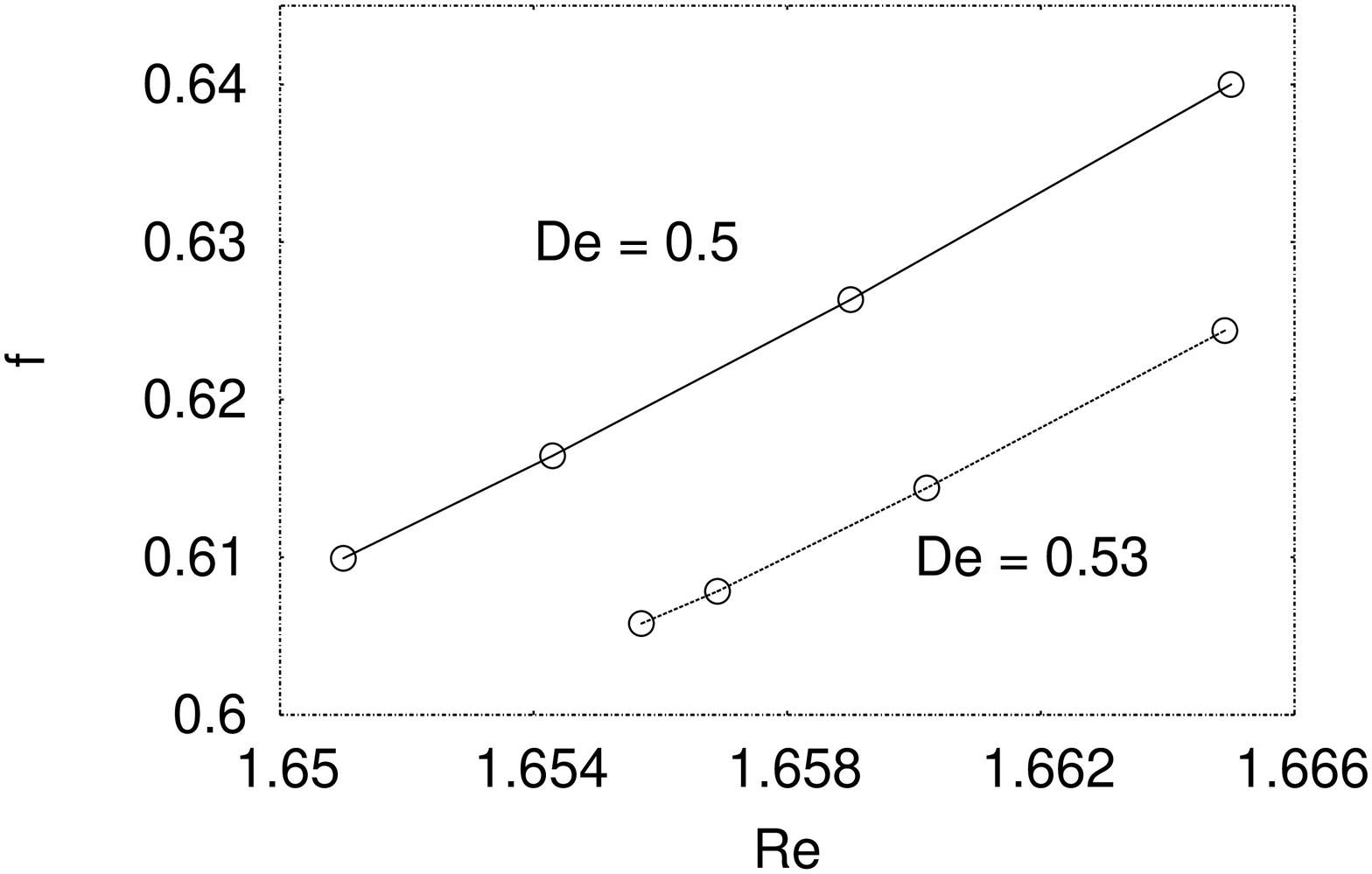}
\includegraphics[angle =0, width =0.45\textwidth]{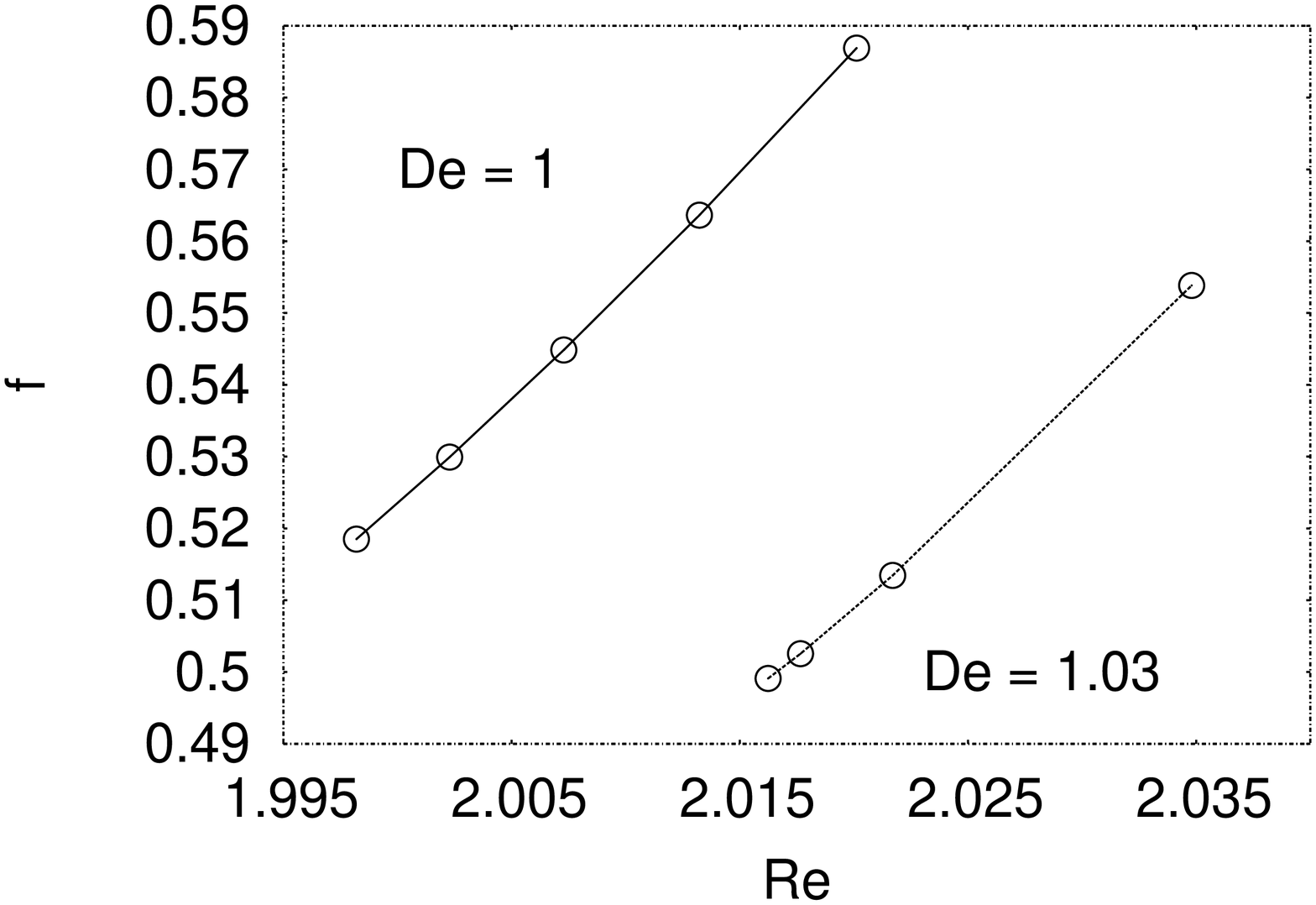} 
\includegraphics[angle =0, width =0.45\textwidth]{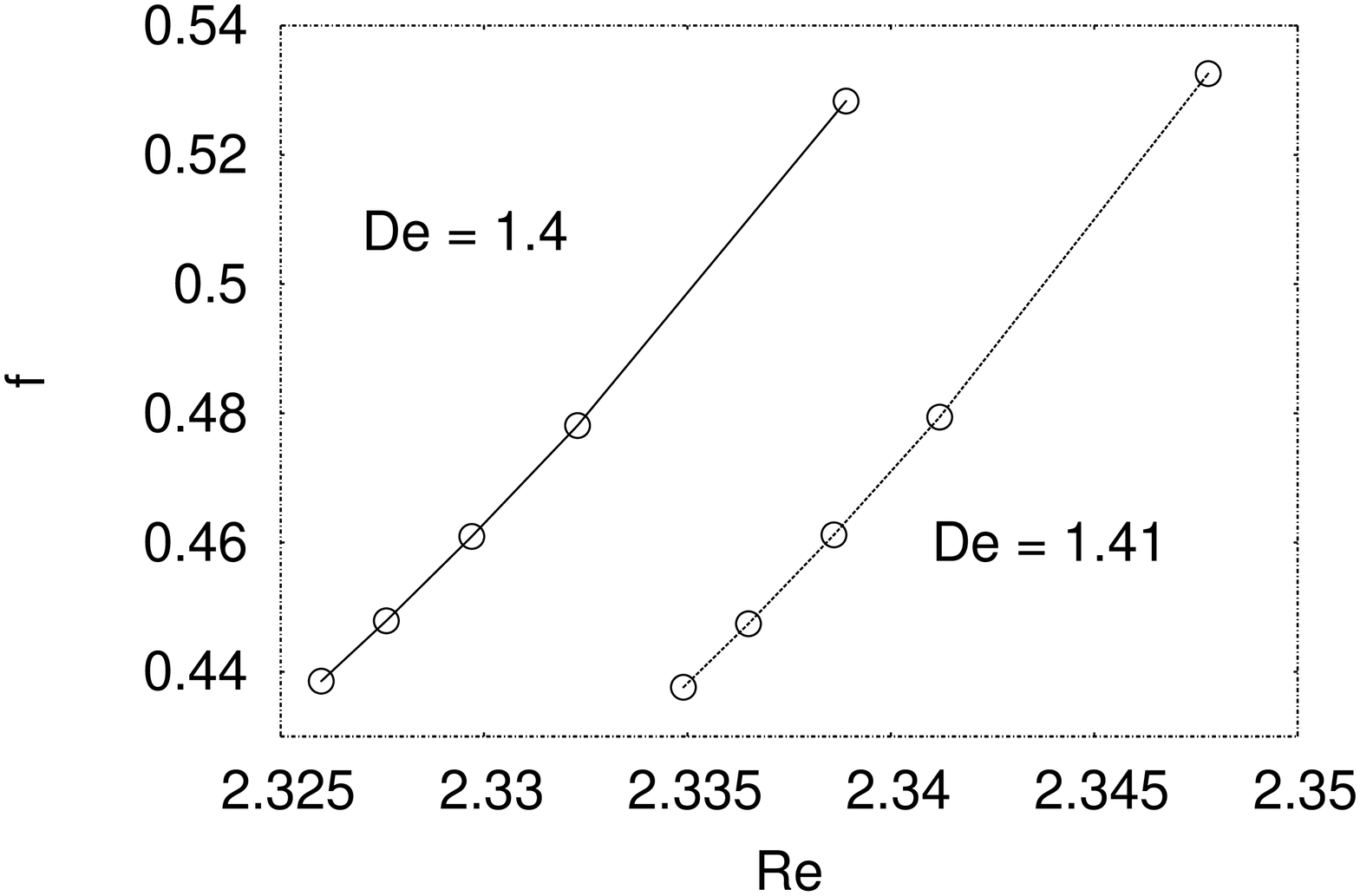} 
\caption{CH simulations at fixed $\De$. the
drag coefficient $f$ is found to increase with the Reynolds
number. The comparison between the various curves shows that the drag
coefficient reduces with the Deborah number.}
\label{fig:DR_decon}
\end{figure}

\medskip

Numerical simulations have been performed to check the consistency of
these results and their outcome is summarized in
figure~\ref{fig:DR_decon}. Here, the Deborah number has been fixed at
different values and the drag coefficient has been plotted versus the
Reynolds number. While $f$ increases with $\Rey$, as expected, larger
Deborah numbers are always found to be associated to smaller drag
coefficients. 

\section{Conclusions}

The weakly nonlinear dynamics of a viscoelastic Kolmogorov flow has
been analysed both analytically and numerically. The physical reasons
for considering this flow are that, despite the fundamental difference
consisting in the absence of material boundaries, it has several
analogies with channel flows and is one of the few well-known exact
solutions of the Oldroyd-B model.

The linear stability analysis for the Kolmogorov flow had already been
developed by \cite{B05}. No insights had however been given for the
weakly nonlinear stage of evolution. This regime amounts to
considering values of the Reynolds number close to the marginal
stability curve separating stable from unstable regions of the
phase-space.  \\ In the general nonlinear case (\ie~for arbitrarily
large distances from the marginal curve), there is no way to solve the
fully nonlinear equations. Conversely, close to the marginal curve,
asymptotic perturbation techniques can be employed to capture the
weakly nonlinear dynamics.

We found that the weakly nonlinear dynamics is
described by Cahn--Hilliard-like equations, with coefficients dependent
on the Deborah number.  The behaviour of these coefficients with
respect to $\De$ reveals that there exists a critical value of the
Deborah number, where the system bifurcates to another regime. The
resulting nonlinear equation still has a Cahn--Hilliard form, but
contains a novel, fifth-order nonlinearity.

Above the critical $\De$, the ``hydrodynamic'' kink-antikink
structures are replaced by generalized structures. We have shown that
their processes of annihilation are slowed down with respect to the
standard Cahn-Hilliard equation.  We also found a purely nonlinear,
subcritical mechanism of instability, which occurs for sufficiently
large amplitudes of the initial perturbation.

Our results demonstrate that, for hydrodynamical systems governed by a
standard Cahn--Hilliard equation, the presence of an additional
parameter might lead to higher-order nonlinear dynamics. A system where
similar phenomena are to be expected is the stratified Kolmogorov flow
investigated by \cite{BY05}, with the role of elasticity played by
stratification.

Our results have been obtained both exploiting the multiscale
expansion and via direct numerical simulations of the original
equations and their coarse-grained version. The agreement between the
Cahn--Hilliard dynamics and the full-resolved one is excellent even at
large times. This is true for both the standard Cahn--Hilliard and the
generalized one. The asymptotic analysis is thus able to capture all
of the relevant features of the flow.

In the last part of the work, we have presented some consequences for
the problem of drag reduction. Although it is not common to talk about
this effect in non-turbulent flows, we have shown that, even in the
weakly nonlinear case, the injection of polymers induces a reduction
in the drag coefficient, via the stabilization of the basic
flow. Using the results of the nonlinear analysis, we have been able
to give an analytical expression for the flow enhancement due to the
polymers. The main qualitative conclusion is that drag reduction stems
from a stabilization of the flow and appears to be a phenomenon
coupling large and small scales.

\section*{Acknowledgments}

This work has been supported by the Italian MIUR COFIN 2005 project
n. 2005027808 (GB,AM), by Fondazione CRT-Progetto Lagrange (AB)  and by the
 European  Networks ``Stirring and Mixing'' HPRN-CT2002-00300 (AC) and
 ``Non-ideal turbulence'' HPRN-CT-2000-00162 (MV).  


\newpage
\newpage

\appendix
\section{Squire's theorem for Oldroyd-B}\label{app:squire}
Consider a parallel flow ${\bm U} = (U(z),0)$. To investigate its stability
properties one writes the linearized, nondimensional equations
\begin{eqnarray}
\partial_t {\bm w} + ({\bm u}\cdot{\bm \nabla}) {\bm w}
+ ({\bm w}\cdot{\bm \nabla}) {\bm u}
=-{\nabla q} + \beta {{\Rey}}^{-1} {\Delta} {\bm w} +
\nonumber \\
+ (1-\beta) \,{{\Rey}}^{-1}\,
{{\De}}^{-1}\,
{\bm \nabla}\cdot {\bm \zeta}
\label{eq:8}
\end{eqnarray}
\begin{eqnarray}
\partial_t {\bm \zeta} + ({\bm u}\cdot{\bm \nabla}) {\bm \zeta}
+ ({\bm w}\cdot{\bm \nabla}) {\bm \sigma}
= ({\bm \nabla \bm u})^T \cdot {\bm \zeta}
+ ({\bm \nabla \bm w})^T \cdot {\bm \sigma} + \nonumber \\
+ {\bm \zeta} \cdot ({\bm \nabla \bm u})
+ {\bm \sigma} \cdot ({\bm \nabla \bm w})
-{{\De}}^{-1} {\bm \zeta}
\label{eq:9}
\end{eqnarray}
where ${\bm w}$ is the perturbation of the basic profile
${\bm u}$,
and ${\bm \zeta}$ is the perturbation of the
basic stress tensor ${\bm \sigma}$.\\
We now perform a Fourier transform in the directions $x$ and $y$, and in time,
\begin{equation}
 w_i(x,y,z,t)= \int d\omega\, d k_x\, d k_y\,
e^{-i\omega t+ k_x x + k_y y}\,\hat{w}_i(k_x,k_y,\omega,z)
\end{equation}
\begin{equation}
\zeta_{ij}(x,y,z,t)= \int d\omega\, d k_x\, d k_y\,
e^{-i\omega t+ k_x x + k_y y}\,\hat{\zeta}_{ij}(k_x,k_y,\omega,z)
\end{equation}
Introducing the notation
\begin{equation}
\begin{array}{c}
\begin{array}{ccc}
{\sf k}=\left(\begin{array}{c}k_x\\k_y\end{array}\right) &
{\sf u}= \left(\begin{array}{c}  U(z)\\0\end{array}\right) &
\hat{\sf w}=\left(\begin{array}{c}\hat{w}_x\\ \hat{w}_y\end{array}\right)
\end{array} \\
\begin{array}{cccc}
\hat{\sf t}=
\left(\begin{array}{c}\hat{\zeta}_{xz} \\ \hat{\zeta}_{yz}\end{array}\right)
&
\hat{\sf z}=
\left(\begin{array}{cc}\hat{\zeta}_{xx}&\hat{\zeta}_{xy}\\
\hat{\zeta}_{yx} &\hat{\zeta}_{yy}\end{array}\right)
&
{\sf r}=
\left(\begin{array}{c}{\sigma}_{xz}\\{\sigma}_{yz}\end{array}\right)
&
{\sf s}=
\left(\begin{array}{cc} {\sigma}_{xx}&{\sigma}_{xy}\\
{\sigma}_{yx}&{\sigma}_{yy}\end{array}\right)
\end{array}
\end{array}
\end{equation}
the linearized equations in normal modes
take the form
\begin{eqnarray}
(-i\omega + i\,{\sf k}^T\cdot{\sf u}) \hat{\sf w}
+\hat{w}_z \frac{d {\sf u}}{dz}
=-i{\sf k}\hat{q}+\beta{{\Rey}}^{-1}(-{\sf k}^2+\frac{d^2}{dz^2})\hat{\sf w} +
\nonumber \\
+ (1-\beta) \,{{\Rey}}^{-1}\,
{{\De}}^{-1}\, \left(
i \hat{\sf z}^T \cdot {\sf k}
+ \frac{d}{dz} \hat{\sf t} \right)
\label{eq:10}
\end{eqnarray}
\begin{eqnarray}
(-i\omega + i {\sf k}^T\cdot{\sf u})\hat{w}_z =
-\frac{d\hat{q}}{dz} + \beta {{\Rey}}^{-1} (-{\sf k}^2+\frac{d^2}{dz^2})
\hat{w}_z + \nonumber \\
+ (1-\beta) \,{{\Rey}}^{-1}\,
{{\De}}^{-1}\, \left(
i {\sf k}^T \cdot \hat{\sf t} + \frac{d}{dz} \hat{\zeta}_{zz} \right)
\label{eq:11}
\end{eqnarray}
\begin{eqnarray}
(-i\omega + i {\sf k}^T \cdot {\sf u}+{{\De}}^{-1})\hat{\sf z}
+\hat{w}_z \frac{d}{dz}{\sf s} =
\hat{\sf t}\cdot \frac{d{\sf u}}{dz}^T + \frac{d{\sf u}}{dz}\cdot\hat{\sf t}^T+
\nonumber\\
+i({\sf s}\cdot{\sf k})\hat{\sf w}^T + i\hat{\sf w}({\sf k}^T \cdot {\sf s})+
{\sf r}\frac{d}{dz}\hat{\sf w}^T + \frac{d \hat{\sf w}}{dz} {\sf r}^T
\label{eq:12}
\end{eqnarray}
\begin{eqnarray}
(-i\omega + i {\sf k}^T\cdot{\sf u}+{{\De}}^{-1})\hat{\sf t}
+\hat{w}_z \frac{d}{dz}{\sf r} =
\hat{\zeta}_{zz} \frac{d{\sf u}}{dz}+\nonumber\\
+ i({\sf s}\cdot{\sf k})\hat{w}_z + i \hat{\sf w}({\sf r}^T\cdot{\sf k})
+{\sf r}\frac{d}{dz}\hat{w}_z + \frac{d}{dz}\hat{\sf w}
\label{eq:13}
\end{eqnarray}
\begin{eqnarray}
(-i\omega + i {\sf k}^T\cdot{\sf u}+{{\De}}^{-1})\hat{\zeta}_{zz} =
2i({\sf r}^T\cdot{\sf k})\hat{w}_z + 2\frac{d}{dz}\hat{w}_z
\label{eq:14}
\end{eqnarray}
Consider the following transformation
\begin{equation}
\begin{array}{c}
\begin{array}{cccc}
\overline{k}_x=|{\sf k}|
&
\overline{w}_x=\frac{{\sf k}^T \cdot \hat{\sf w}}{|{\sf k}|}
&
\overline{w}_z = \hat{w}_z
&
\overline{q}=\frac{|{\sf k}|}{k_x}\hat{q}
\end{array}
\vspace{5pt}\\
\begin{array}{ccc}
\overline{{\Rey}}=\frac{k_x}{|{\sf k}|}{\Rey}
&
\overline{{\De}}=\frac{k_x}{|{\sf k}|}{\De}
&
\overline{\omega}=\frac{|{\sf k}|}{k_x}\omega
\end{array}
\vspace{5pt}\\
\begin{array}{ccc}
 \overline{t}_x= \frac{k_x}{|{\sf k}|}\;
\frac{{\sf k}^T \cdot \hat{\sf t}}{|{\sf k}|}
&
\overline{\zeta}_{xx}= \frac{k_x}{|{\sf k}|}\;
\frac{{\sf k}^T \cdot \hat{\sf z} \cdot {\sf k}}{|{\sf k}|^2}
&
\overline{\zeta}_{zz}=\frac{k_x}{|{\sf k}|}\hat{\zeta}_{zz}
\end{array}
\end{array}
\end{equation}
From \eqref{eq:10}-\eqref{eq:14} one can derive the equations
for the overlined variables
\begin{eqnarray}
\left[-i\overline{\omega} + i\,\overline{k}_x U(z)\right] \overline{w}_x
+\overline{w}_z \frac{dU}{dz}
=-i \,\overline{k}_x \overline{q}+
\beta {\overline{{\Rey}}}^{-1}(-\overline{k}_x^2+
\frac{d^2}{dz^2})\overline{w}_x +
\nonumber \\
+ (1-\beta) \,{\overline{{\Rey}}}^{-1}\,
{\overline{{\De}}}^{-1}\, \left(
i \,\overline{k}_{x} \overline{\zeta}_{xx}
+ \frac{d}{dz} {\overline{t}}_x \right)
\label{eq:16}
\end{eqnarray}
\begin{eqnarray}
\left[-i\overline{\omega} + i\,\overline{k}_x U(z)\right] \overline{w}_z =
-\frac{d\hat{q}}{dz} +
\beta \overline{{\Rey}}^{-1} (-\overline{k}_x^2+
\frac{d^2}{dz^2}) \overline{w}_z + \nonumber \\
+ (1-\beta) \,{\overline{{\Rey}}}^{-1}\,
{\overline{{\De}}}^{-1}\, \left(
i \,\overline{k}_x \overline{t}_x + \frac{d}{dz} \overline{\zeta}_{zz} \right)
\label{eq:17}
\end{eqnarray}
\begin{eqnarray}
\left[-i\overline{\omega} + i\overline{k}_x U(z)
+\overline{{\De}}^{-1}\right] \overline{\zeta}_{xx}
+\overline{w}_z \frac{d\overline{s}_{xx}}{dz} =
2 \overline{t}_x \frac{dU}{dz}+
2i\overline{k}_x\overline{s}_{xx}\overline{w}_x +
2 \overline{r}_x \frac{d\overline{w}_x}{dz}
\label{eq:18}
\end{eqnarray}
\begin{eqnarray}
\left[-i\overline{\omega} + i\overline{k}_x U(z)
+\overline{{\De}}^{-1}\right] \overline{t}_x
+\overline{w}_z \frac{d}{dz}\overline{r}_x =
\overline{\zeta}_{zz} \frac{dU}{dz}
+ i\overline{s}_{xx} \overline{k}_x \overline{w}_z +
i \overline{k}_x \overline{w}_x \overline{r}_x \nonumber \\
+\overline{r}_x\frac{d\overline{w}_z}{dz} + \frac{d\overline{w}_x}{dz}
\label{eq:19}
\end{eqnarray}
\begin{eqnarray}
\left[-i\overline{\omega} + i\overline{k}_x U(z)
+\overline{{\De}}^{-1}\right]\overline{\zeta}_{zz}=
2i \overline{k}_x \overline{r}_x \overline{w}_z + 2 \frac{d\overline{w}_z}{dz}
\label{eq:20}
\end{eqnarray}
where we introduced the quantities
\begin{eqnarray}
\overline{s}_{xx} = \frac{{\sf k}^T \cdot {\sf s} \cdot {\sf k}}{|{\sf k}|^2}
= 1+\overline{{\De}}^2 [U'(z)]^2, & &
\overline{r}_x = \frac{{\sf k}^T \cdot {\sf r}}{|{\sf k}|} =
\overline{{\De}}^2 U'(z)
\end{eqnarray}
Equations~\eqref{eq:16}-\eqref{eq:20} are exactly the same as
\eqref{eq:10}-\eqref{eq:14} but with $k_y=0$, $\hat{w}_y=0$,$\zeta_{xy}=
\zeta_{yy}=\zeta_{yz}=0$. Therefore they describe a two-dimensional linear
disturbance of the basic flow at smaller Reynolds and Deborah numbers.
If the three-dimensional perturbation ${\bm w},{\bm \zeta}$ is
unstable at $({\Rey},{\De})$, then the two-dimensional
disturbance $\overline{\bm w},\overline{\bm \zeta}$ is unstable
at $(\overline{{\Rey}},\overline{{\De}})$ and its rate of
growth is larger ($\mathrm{Im}(\overline{\omega})\ge\mathrm{Im}(\omega)>0$).


\section{Painlev\'e analysis}\label{app:painleve}
We perform a Painlev\'e analysis to ascertain whether the fifth-order
equation \eqref{eq:Cahn-Hilliard-Gen} is integrable as the usual cubic
Cahn--Hilliard equation \eqref{eq:Cahn-Hilliard}.

After rescaling dependent and independent variables, the
stationary equation takes the form:
\begin{equation}\label{eq:pain1}
-u-\frac{u^3}{3}  - \lambda \partial_x^2 u + \frac{\gamma}{5} u^5 = 0 \, .
\end{equation}
The Painlev\'e test consists in checking whether the structure of the
solution around singularities in the complex plane has the form of a
Laurent series. A simple balance of the last two terms in the equation
indicates that the singularity has order $-1/2$. The putative Laurent
series should then be sought as:
\begin{equation}\label{eq:laurent}
u(z)=z^{-1/2}\left[u_0 + u_1 z + u_2 z^2 + u_3 z^3 + \cdots\right] \,
,
\end{equation}
where $z$ is the complex variable denoting the separation from the
singularity $z_*$. When the series \eqref{eq:laurent} is inserted into
equation \eqref{eq:pain1}, a hierarchy of equations of the form $a_k
u_k = b_k$ is obtained. The $a_k$'s and $b_k$'s can be calculated in
terms of $u_{k-1}, \cdots, u_0$. The impossibility for an arbitrary
equation to have a Laurent series expansion is due to resonances, \ie
values of $k$ such that $a_k=0$. Integrability is equivalent to
checking that $b_k=0$ for the orders corresponding to resonances. In
our case, it is easy to check that
\begin{equation}\label{eq:pain2}
a_k = - \lambda \left(k-\frac{1}{2}\right)\left(k-\frac{3}{2}\right) +
\gamma u_0^4 \,; \qquad u_0=\left(\frac{15 \lambda}{4
\gamma}\right)^{1/4} \mapsto a_k=-\lambda(k+1)(k-3)\, .
\end{equation}
The resonance is therefore at the third order and we need to perform
the explicit calculation up to that order to check whether or not
$b_3=0$. The algebra is elementary and the coefficients are:
\begin{equation}\label{eq:pain3}
u_1 = \frac{u_0^3}{12 \lambda}, \, u_2=\frac{u_0}{\lambda}\left[
\frac{1}{3} + \frac{5}{128 \gamma}\right] \, .
\end{equation}
Using these values, one can verify that
\begin{equation}\label{eq:pain4}
b_3 = 2 \gamma u_0^2 u_1^3 + 4 \gamma u_0^3 u_1 u_2  - u_1 - u_0^2u_2 -u_0u_1^2
\end{equation}
vanishes and the Painlev\'e test is satisfied.


\begin{thebibliography}{99}
\bibitem[Ablowitz \& Clarkson(1991)]{AC91}
  {\sc Ablowitz, M.J. \& Clarkson, P.A.} 1991 {\it Solitons, Nonlinear Evolution Equations and Inverse Scattering.} Cambridge University Press.
\bibitem[Arnold \& Meshalkin(1960)]{AM60}
  {\sc Arnold, V.I. \& Meshalkin, L.} 1960 A.N. Kolmogorov's seminar on
  selected problems of analysis (1958--1959). {\it Uspekhi Mat. Nauk} {\bf
  15}, 247--250.

\bibitem[Balmforth \& Young(2005)]{BY05}
  {\sc Balmforth, N.J., Young, Y.-N.} 2995 Stratified Kolmogorov flow. Part
  2. {\it J. Fluid Mech.} {\bf 528}, 23-42. 


\bibitem[Bayly \etal(1988)]{BOH88} 
  {\sc Bayly, B.J., Orszag, S.A., Herber,T.} 1988 Instability mechanisms in
  shear-flow transition. {\it Annu. Rev. Fluid Mech.} \textbf{20}, 359-391.



\bibitem[Bensoussan \etal(1978)]{BLP78}
  {\sc Bensoussan, A., Lions, J.--L. \& Papanicolau, G.} 1978 {\it Asymptotic Analysis for Periodic Structures.} North--Holland.
\bibitem[Bird \etal(1987)]{B87}
  {\sc Bird, R.B., Hassager, O., Armstrong, R.C. \& Curtiss, C.F.} 1987 {\it Dynamics of Polymeric Liquids.} Wiley-Interscience.
\bibitem[Boffetta \etal(2005a)]{B05}
  {\sc Boffetta, G., Celani, A., Mazzino, A., Puliafito, A. \& Vergassola, M.} 2005a The viscoelastic Kolmogorov flow: eddy viscosity and linear stability. {\it J. Fluid Mech.} {\bf 523}, 161--170.
\bibitem[Boffetta \etal(2005b)]{BCM05}
  {\sc Boffetta, G., Celani, A. \& Mazzino, A.} 2005b Drag reduction in the turbulent Kolmogorov flow. {\it Phys. Rev. E} {\bf 71}, 036307.



\bibitem[Bray(2002)]{B02} 
  {\sc Bray, A.J.} 2002 Theory of phase-ordering kinetics. {\it Advances in Physics}  {\bf 51}, 481587



\bibitem[Cahn \& Hilliard(1958)]{CH58}
  {\sc Cahn, J.W. \& Hilliard, J.E.} 1958 Free energy of a non uniform system. {\it J. Chem. Phys.} {\bf 28}, 258.


\bibitem[Canuto \etal(1988)]{CHQZ88}
  {\sc Canuto, C., Hussaini, M.Y., Quarteroni, A. \& Zang, T. A.} 1988 {\it
  Spectral Methods in Fluid Dynamics.} Springer-Verlag.










\bibitem[Gama \etal (1994)]{GVF94}
  {\sc Gama, S., Vergassola, M. \& Frisch, U.} 1994 
Negative eddy-viscosity in isotropically forced two-dimensional flow:
linear and nonlinear dynamics. {\it J. Fluid Mech.} {\bf 260}, 95--126.

\bibitem[Geraschenko \etal (2005)]{GCS05}
  {\sc Geraschenko, S., Chevallard, C. \& Steinberg, V.} 2005 
Single-polymer dynamics: Coil-stretch transition in a random flow.
 {\it Europhys. Lett.} {\bf 71}(2), 221--227.


\bibitem[Hinch(1977)]{H77}
  {\sc Hinch, E.J.} 1977 Mechanical models of dilute polymer solutions in strong flows. {\it Phys. Fluids} {\bf 20}, S22--S30.

\bibitem[Kawasaki \& Ohta(1982)]{KO82}
  { \sc Kawasaki, K. \& Ohta, T} 1982 Kink dynamics in one-dimensional
  nonlinear systems. {\it Physica A} {\bf 116} 573.


\bibitem[Karttunen \etal(2004)]{KVLN04} 
  {\sc Karttunen, M., Vattulainen, I. \& Lukkarinen, A.} 2004 {\it Novel
    Methods in Soft Matter Simulations.} Springer-Verlag.

\bibitem[Khouider \etal(2003)]{KMK03} 
  {\sc Khouider, B., Majda, A.J. \& Katsoulakis, M.A.} 2003 Coarse-grained
  stochastic models for tropical convection and climate. {\it Proc. Natl. Acad. Sci} {\bf 100}, 1194111946.




\bibitem[Kowalesvki(1889)]{K889}
  {\sc Kowalesvki, S.} 1889 Sur le probl\`eme de la rotation d'un corps solide autur d'un point fixe. {\it Acta Math.} {\bf 12}, 177--232.
\bibitem[Kowalesvki(1890)]{K890}
  {\sc Kowalesvki, S.} 1890 Sur une propri\'et\'e du syst\`eme d'\'equations diff\'erentielles qui d\'efinit la rotation d'un corpe solide autour d'un point fixe. {\it Acta Math.} {\bf 14}, 81--93.
\bibitem[Larson(1992)]{L92}
  {\sc Larson, R.G.} 1992 Instabilities in viscoelastic flows. {\it Rheol. Acta} {\bf 31}, 213--263.
\bibitem[Legras \& Villone(2003)]{LV03}
  {\sc Legras, B. \& Villone, B.} 2003 Dispersive and friction-induced stabilization of the Cahn--Hilliard inverse cascade. {\it Physica D} {\bf 175}, 139--166.
\bibitem[Lumley(1969)]{L69}
  {\sc Lumley, J.} 1969 Drag reduction by additives. {\it Annu. Rev. Fluid Mech.} {\bf 1}, 367--384.

\bibitem[Manfroi \& Young(1999)]{MY99}
  {\sc Manfroi, A., \& Young, W.} 1999 Slow evolution of zonal jets on the
  beta-plane. {\it J. Atmos. Science} {\bf 56}, 784--800.

\bibitem[Meshalkin \& Sinai(1961)]{MS61}
  {\sc Meshalkin, L. \& Sinai, Ya G.} 1961 Investigation of the stability of a stationary solution of a system of equations for the plane movement of an incompressible viscous fluid. {\it J. Appl. Math. Mech.} {\bf 25}, 1700--1705.
\bibitem[Nadolink \& Haigh(1995)]{NH95}
  {\sc Nadolink, R.H. \& Haigh, W.W.} 1997  Bibliography on skin friction
  reduction with polymers and other boundary-layer additives. {\it ASME
  Appl. Mech. Rev.} {\bf 48}, 351--460.


\bibitem[Nepomnyashchyi(1976)]{N76}
  {\sc Nepomnyashchyi, A.A.} 1976 On the stability of the secondary flow of a
  viscous fluid in an infinite domain. {\it Appl. Math. Mech.} {\bf 40}, 886--891.


\bibitem[Oldroyd(1950)]{O50}
  {\sc Oldroyd, J.G.} 1950 On the formulation of rheological equations of state. {\it Proc. Roy. Soc. London Ser A} {\bf 200}, 523--541.
\bibitem[Painlev\'e(1897)]{P897}
  {\sc Painlev\'e, P.} 1897 {\it Le\c cons sur la Th\'eorie Analytique des \'Equations Differentielles.} Hermann, Paris.

\bibitem[Pedlosky(1987)]{P87}
  {\sc Pedlosky, J.} 1987 {\it Geophysical fluid dynamics} 2nd edition. Springer.


\bibitem[She(1987)]{She87}
  {\sc She, Z.S.} 1987 Metastability and vortex pairing in the Kolmogorov flow. {\it Phys. Lett. A} {\bf 124}, 161--164.
\bibitem[Sivashinsky(1985)]{S85}
  {\sc Sivashinsky, G.I.} 1985 Weak turbulence in periodic flows. {\it Physica D} {\bf 17}, 243--255.
\bibitem[Sreenivasan \& White(2000)]{SW00}
  {\sc Sreenivasan, K.R. \& White, C.M.} 2000 The onset of drag reduction by dilute polymer additives, and the maximum drag reduction asymptote. {\it J. Fluid Mech.} {\bf 409}, 149--164.
\bibitem[Sureshkumar \etal(1997)]{SBH97}
  {\sc Sureshkumar, R., Beris, A.N., \& Handler, R.A.} 1997  Direct numerical simulation of polymer-induced drag reduction in turbulent channel flow. {\it Phys. Fluids} {\bf 9}, 743--755.
\bibitem[Toms(1949)]{T49}
  {\sc Toms, B.A.} 1949 Observation on the flow of linear polymer solutions through straight tubes at large
  Reynolds numbers. {\it Proc. 1st International Congress on Rheology} {\bf 2}, 135--141.


\bibitem[Vattulainen \& Karttunen(2005)]{VK05} 
  {\sc Vattulainen, I. \& Karttunen, M.} 2005 {\it Computational Nanotechnology},
    edited by M. Rieth and W. Schommers. American Scientific Press, in press.

\bibitem[Virk(1975)]{V75}
  {\sc Virk, P. S.} 1975 Drag reduction fundamentals. {\it AIChE Journal} {\bf 21}, 625--656.


\end{thebibliography}
\end{document}